\documentclass[aps,prb,showpacs,preprintnumbers,amsmath,amssymb,twocolumn,floatfix,footinbib]{revtex4-2}

\usepackage{amsmath,amssymb}
\usepackage{graphicx}
\usepackage{bm}
\usepackage{bbm}
\usepackage{color}
\usepackage{natbib}
\usepackage{hyperref}
\hypersetup{
  colorlinks,
  citecolor=blue,
  linkcolor=red,
  urlcolor=blue}
\usepackage{times}
\usepackage[T1]{fontenc}

\usepackage{multirow}
\usepackage{siunitx}
\usepackage{placeins}

\newcommand{\cfig}[1]{Fig.~\ref{#1}}
\newcommand{\csec}[1]{Sec.~\ref{#1}}
\newcommand{\ceqn}[1]{Eq.~(\ref{#1})}
\newcommand{\capp}[1]{App.~\ref{#1}}
\newcommand{\ctab}[1]{Tab.~\ref{#1}}
\newcommand{\cref}[1]{Ref.~\onlinecite{#1}}

\newcommand{\pd}{\phantom{\dagger}}

\newcolumntype{C}[1]{>{\centering\arraybackslash}m{#1}}
\newcolumntype{L}[1]{>{\raggedright\arraybackslash}m{#1}}
\newcolumntype{R}[1]{>{\raggedleft\arraybackslash}m{#1}}

\newcounter{mylabelcounter}
\makeatletter
\newcommand{\labelText}[2]{%
	#1\refstepcounter{mylabelcounter}%
	\immediate\write\@auxout{%
		\string\newlabel{#2}{{1}{\thepage}{{\unexpanded{#1}}}{mylabelcounter.\number\value{mylabelcounter}}{}}%
	}%
}
\makeatother

\begin{document}
\title{Interplay of Fractional Chern Insulator and Charge-Density-Wave Phases\\ in Twisted Bilayer Graphene}

\author{Patrick Wilhelm}
\email{patrick.wilhelm@uibk.ac.at}
\affiliation{Institut f\"ur Theoretische Physik, Universit\"at Innsbruck, A-6020 Innsbruck, Austria}
\author{Thomas C. Lang}
\affiliation{Institut f\"ur Theoretische Physik, Universit\"at Innsbruck, A-6020 Innsbruck, Austria}
\author{Andreas M. L\"{a}uchli}
\affiliation{Institut f\"ur Theoretische Physik, Universit\"at Innsbruck, A-6020 Innsbruck, Austria}

\begin{abstract}
	We perform an extensive exact diagonalization study of interaction driven insulators in spin- and valley-polarized moir\'{e} flat bands of twisted bilayer graphene aligned with its hexagonal boron nitride substrate. In addition to previously reported fractional Chern insulator phases, we provide compelling evidence for competing charge-density-wave phases at multiple fractional fillings of a realistic single-band model. A thorough analysis at different interlayer hopping parameters, motivated by experimental variability, and the role of kinetic energy at various Coulomb interaction strengths highlight the competition between these phases. The interplay of the single-particle and the interaction induced hole dispersion with the inherent Berry curvature of the Chern bands is intuitively understood to be the driving mechanism for the ground-state selection. The resulting phase diagram features remarkable agreement with experimental findings in a related moir\'{e} heterostructure and affirms the relevance of our results beyond the scope of graphene based materials.

\end{abstract}
\date{\today}

\maketitle

\section{Introduction} 

Over the course of the past three years, twisted bilayer graphene (TBLG) and related moir\'{e} heterostructures emerged as promising platforms for the study of interaction effects in realistic flat band systems. The ability to engineer bands of minimal bandwidth via two stacked graphene sheets subject to a relative \emph{magic} twist-angle of about $1.1^\circ$, in combination with the excellent experimental tunability of the band filling through electric gates has lead to a tremendous growth of interest in the field of graphene-based moir\'{e} materials. Experimental observations of correlated insulators in proximity to potentially unconventional superconductivity \cite{Cao2018, Cao2018a, Yankowitz2019, Lu2019} raised hopes that the study of this composite system may shine light on the long-standing mystery of the mechanism behind high-temperature superconductivity in cuprates. More recent experiments point to the possibility that these correlated insulators and superconductivity might have distinct microscopic origins though \cite{Stepanov2020, Saito2020, Liu2020c, Arora2020}. The nature of the superconducting phase and its pairing mechanism is generally subject to hot debates, including exotic proposals involving topological solitons -- skyrmions -- carrying charge $2e$ \cite{Christos2020, Scheurer2020, Khalaf2020, Chatterjee2020, Hu2019, Julku2020, Xie2020b}. Further experimental signatures include ferromagnetism \cite{Sharpe2019} and a quantized anomalous Hall effect \cite{Serlin2019} in TBLG aligned with the hexagonal boron nitride (hBN) substrate (TBLG/hBN), which suggests that a Chern insulator may be realized in TBLG related materials. The incorporation of interactions naturally leads to the question whether a fractional Chern insulator may form in TBLG, which has been answered affirmatively using exact diagonalizations in Refs.~\cite{Abouelkomsan2020,Repellin2020} and analytically in Ref.~\cite{Ledwith2020}. 
\begin{figure}[t]
	\includegraphics[width=0.9\linewidth]{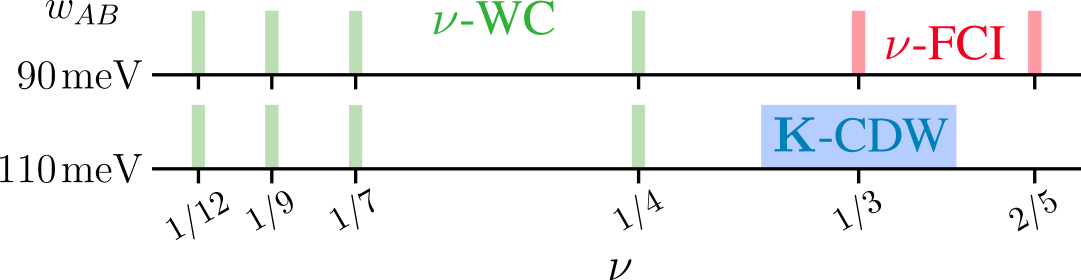}
	\caption{Schematic TBLG/hBN phase diagram of the identified order tendencies over their observed band filling range $\nu$ for the studied hopping parameters ${w_{AB}=\SI{90}{meV}}$ and ${w_{AB}=\SI{110}{meV}}$ in the strongly interacting regime. The classification is based on the results compiled in \csec{sec:composition}.
		We find two fractional Chern insulators (FCI), a series of Wigner crystals (WC) locked at specific fillings, as well as a charge-density-wave (CDW) phase around $1/3$ filling, with a seemingly finite density range extent. \label{fig:phase_diagram}
	}
\end{figure}
Studies of TBLG-inspired Hofstadter models on the honeycomb lattice reiterate the importance of fractional quantum Hall (FQH) states at fillings ${\nu=1/3}$ as well as ${\nu=2/5}$ \cite{Andrews2020}. Further exact diagonalization \cite{Xie2020a} and DMRG based \cite{Soejima2020, Kang2020} calculations support the formation of a Chern insulator as well as the possibility for different types of spatial symmetry breaking charge-density-waves (CDW) in pure TBLG. The former predictions are corroborated by the experimental observation of interaction induced Chern insulators at multiple integer fillings \cite{Nuckolls2020}.

Very recently, novel sensing techniques were used to reveal insulating behavior at fractional single-band fillings ${\nu=1/2}$, $2/3$, $2/5$, $1/3$, $1/4$, $1/7$ of a related moir\'{e} heterostructure based on transition metal dichalcogenides (TMD), such as $\text{WS}_2$/$\text{WSe}_2$ \cite{Xu2020, Regan2020}. The order mediated by Coulomb interactions is suggested to be of CDW-type, realizing generalized Wigner crystals (WC) that are locked to certain commensurate filling fractions of the moir\'{e} lattice and spontaneously break translational symmetry. This is in accordance with the possibility to engineer flat bands and the resulting signatures of collective phases reported in \cref{Wang2020} for twisted bilayers of $\text{WSe}_2$ close to half-band filling.

The intrinsic competition of FQH states with WCs and CDWs at fractional fillings has a long history and dates back to early studies of interaction effects in the two-dimensional electron gas subject to a magnetic field \cite{Fukuyama1979, Yoshioka1983a,Yoshioka1983, Moessner1996, Rezayi1999, Haldane2000, Yang2001}. More recently the lattice generalization of a topological Laughlin-like state, the fractional Chern insulator (FCI), has attracted considerable interest \cite{Neupert2011, Tang2011, Sun2011, Regnault2011,Laeuchli2013,Bergholtz2013}. In graphene related systems, a CDW as well as the FCI have been observed experimentally \cite{Rahnejat2011,Spanton2018}. Both types of bulk insulating phases inherently rely on the presence of strong electron-electron interactions, while the FCI additionally requires an effective magnetic field, quantified by a finite Chern number of the fractionally filled band. As the kinetic energy typically weakens such order tendencies, realizations of (nearly) flat bands are typically expected to be a prerequisite to study the competition of these two strongly correlated phases. 

In our work, using large-scale exact diagonalizations, we carefully explore to what extent a similar competition is at work in a realistic spin- and valley-polarized single-band model for TBLG/hBN. We demonstrate that indeed charge ordered states are strong contenders for the ground-state at several fractional fillings, including cases where previous work highlighted the presence of an FCI state \cite{Abouelkomsan2020, Repellin2020}. Furthermore, the nature of the CDWs is shown to go beyond the simple WC-type, realizing stable $\mathbf{K}$-CDW order across a whole range of fillings for suitable band parameters. We show that the quantum geometry, manifest in the inhomogeneous distribution of the Berry curvature, but also the nontrivial momentum dependence of the single-particle dispersion have a strong influence on the FCI/CDW competition beyond the mere presence of a flat Chern band. This understanding allowed us to uncover another FCI state at ${\nu=2/5}$, akin to the results of \cref{Repellin2020}. The acquired intuition in conjunction with the extensive amount of numerical evidence is condensed in the tentative phase diagram of \cfig{fig:phase_diagram}.
Drawing connections to the experiment, the agreement of our results in \cfig{fig:phase_diagram} at ${w_{AB}=\SI{110}{meV}}$ for TBLG/hBN with those of the TMD moir\'{e} system in \cref{Xu2020} suggests a substantial degree of similarity for the physics at play. The added twist of topological nontriviality in TBLG/hBN, however, enables more exotic correlated phases for different band parameters, ensuring again the diversity of physics contained in graphene based moir\'{e} structures. 
 
This work is organized as follows: Section \ref{sec:model} introduces the single-particle model as well as the single-band-projected many-body Hamiltonian and gives an overview of crucial quantities that characterize the model for a certain choice of hopping parameters. We subsequently give a brief overview of the applied numerical method as well as important observables that characterize the discussed correlated phases in \csec{sec:method}. The main volume of numerical results is presented throughout \csec{sec:nu_canonical}, including solid evidence for CDW/WCs at multiple filling fractions as well as the identification of two hierarchy FCI states at ${\nu=1/3}$ and ${\nu=2/5}$. In \csec{sec:composition}, the results are condensed into a tentative phase diagram as a function of the electron filling and important aspects of the phases' nature and stability towards the removal or addition of additional electrons are revealed. This section also demonstrates the commonalities and differences of the two distinguished hopping parameter regions at a glance and allows us to draw possible connections to the experiment in \cref{Xu2020}.

\section{Model \label{sec:model}} 

\begin{figure*}
	\includegraphics[width=0.95\textwidth]{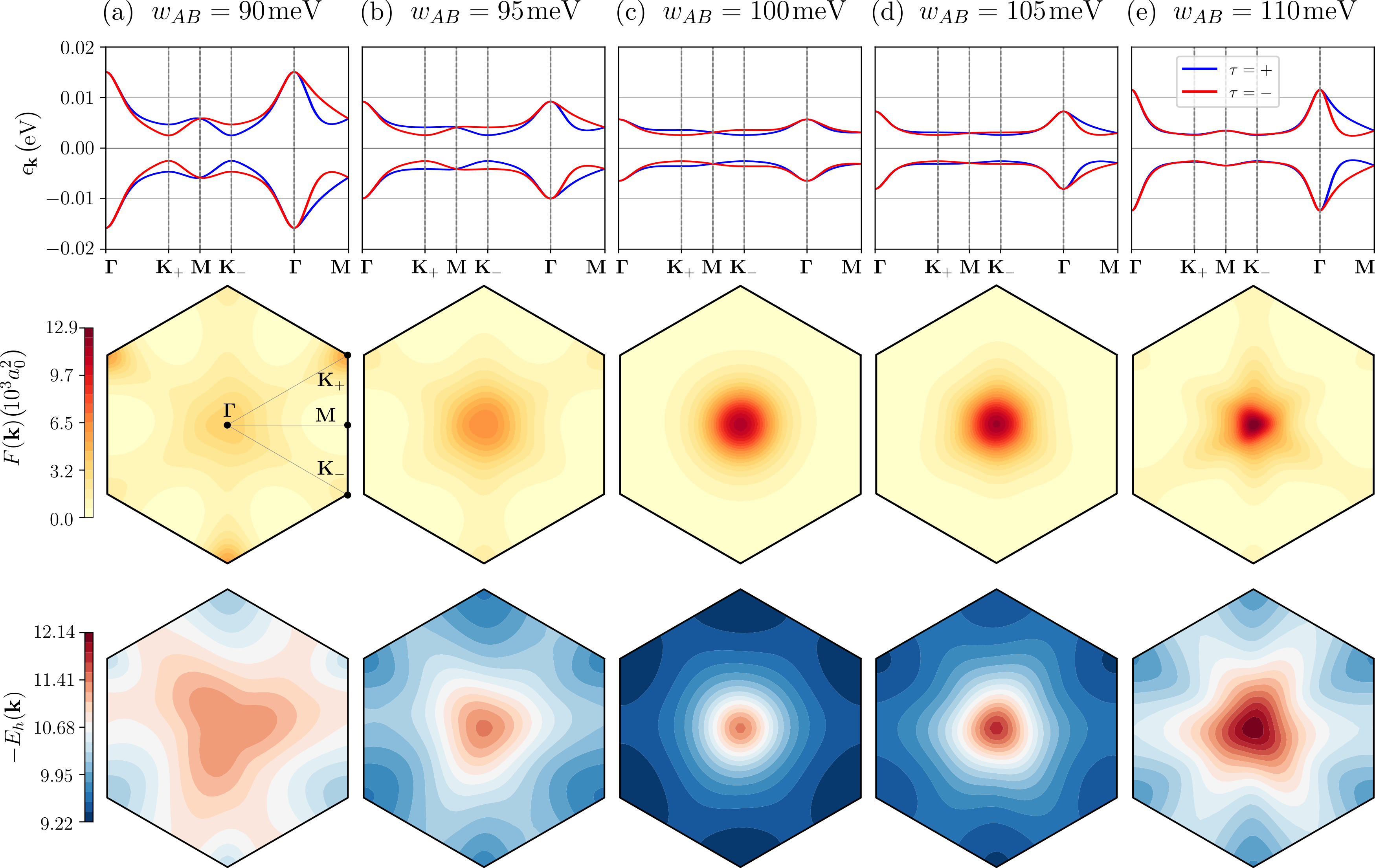}
	\caption{Overview of the single-particle band structure $\epsilon_\mathbf{k}$ (top), the Berry curvature $F(\mathbf{k})$ (middle) and the interaction induced hole dispersion $E_h(\mathbf{k})$ (bottom) of the ${\tau=-}$ valence band for various ${w_{AB}=\SIrange{90}{110}{meV}}$, organized into columns (a) to (e). More remote bands are separated by energetic gaps from the flat band cluster and lie outside the chosen energy window. Common to all $w_{AB}$ is the minimum (maximum) of the valence (conduction) band dispersion as well as the maximum of $-E_h(\mathbf{k})$ at $\mathbf{\Gamma}$. $F(\mathbf{k})$ is redistributed from a relatively uniform case in (a) to a sharp peak at $\mathbf{\Gamma}$ in (e).
	\label{fig:overview}}
\end{figure*}
Our exact diagonalization study is based on the continuum model description of TBLG \cite{Santos2007,Bistritzer2011} at ${\theta = 1.05^\circ}$ (the $(31,1)$ commensurate superlattice in \cref{Santos2012}). We choose the nearest-neighbor hopping amplitude ${t=\SI{2.62}{eV}}$ from graphene and additionally include a phenomenological layer corrugation by using distinct intra- and intersublattice hopping amplitudes $w_{AA}$ and $w_{AB}$ \cite{Koshino2018,Abouelkomsan2020,Bultinck2020a}. Throughout this manuscript we fix ${w_{AA}/w_{AB}=0.7}$, and $w_{AB}$ is varied between the realistic values of \SI{90}{meV} and \SI{110}{meV} to account for model variations and the presence of strain or pressure in samples \cite{Bistritzer2011,Moon2013,Koshino2015,Koshino2018,Abouelkomsan2020}. We assume alignment with the hBN substrate, which, to lowest order, introduces a staggered potential that breaks $C_2$ sublattice symmetry and thus gaps the previously massless Dirac cones at the corners $\mathbf{K}_{\pm}$ of the moir\'{e} Brillouin zone (MBZ)  \cite{Zhang2019a}. The resulting flat valence (conduction) bands of the $\tau=\pm$ valleys, presented in the top row of \cfig{fig:overview}, then acquire a Chern number ${C=\mp 1}$~${(\pm 1)}$. For a realistic substrate induced potential of $\Delta_{\text{hBN}}=\SI{15}{meV}$ \cite{Zhang2019a}, the valence and conduction bands are well separated, such that they may be treated separately for appropriate bandwidths and interaction strengths. Except for an inversion of the valley resolved bands along the $\mathbf{\Gamma}$-$\mathbf{M}$ path and an increased asymmetry of the gaps at $\mathbf{K}_{\pm}$ for lower $w_{AB}$, the single-particle dispersions in \cfig{fig:overview} are qualitatively similar for all considered values of $w_{AB}$. 
A more profound distinction is present in the Berry curvature of the valence ${\tau=-}$ band in the middle row of \cfig{fig:overview}. The rather uniform distribution for ${w_{AB}=\SI{90}{meV}}$ gradually develops a peak at $\mathbf{\Gamma}$ upon increasing $w_{AB}$ to $\SI{110}{meV}$. The analysis remains valid upon switching valley or band, as the Berry curvature is almost identical up to a sign flip, or combined sign coordinate inversion. Minor quantitative differences are the consequence of the slight particle-hole asymmetry of the dispersion. At this stage, it should be noted that the flatness of the Berry curvature is controlled by the combined choice of band parameters and twist angle, enabling a relatively uniform distribution also for ${w_{AB}=\SI{110}{meV}}$ at $\theta=1.15^\circ$ in \cref{Repellin2020}.

We incorporate the Coulomb interaction via the two-dimensional Fourier transform of a Yukawa potential ${V(\mathbf{q}) = (e^2/4 \pi \epsilon \epsilon_0 \Omega) (2 \pi /\sqrt{|\mathbf{q}|^2 + 1/\lambda^2})}$. Here, $e$ and $\epsilon_0$ are the elementary charge and vacuum dielectric constant, respectively, $\Omega$ denotes the total area of the system and $\lambda$ the screening length. The relative dielectric permittivity $\epsilon$ effectively scales the interaction strength, however it is replaced in our treatment by a convex combination of the kinetic and interacting parts of the full Hamiltonian and is thus set to a sensible value of ${\epsilon=2.675}$. If not mentioned otherwise, in accordance with previous authors we choose ${\lambda=L^{\text{M}} \approx \SI{13.4}{nm}}$ to match the moir\'{e} period \cite{Abouelkomsan2020,Liu2020a,Wu2020}. Motivated by experimental signatures \cite{Chen2020, Liu2020b, Sharpe2019, Zondiner2020} and theoretical findings \cite{Bultinck2020a, Repellin2020} we assume full flavor polarization, resulting in an interaction Hamiltonian $H_{\text{int}}$ that acts on spinless fermions of a single valley. Because the two valley flavors of the model are related by time-reversal symmetry, we choose to study electrons in the ${\tau=-}$ valence band with Chern number $C=1$ at an electron filling $\nu$. To incorporate this truncation of the band and flavor interaction channels in the model, we have to project the ordinary density-density interaction operator to the band basis. This step has been detailed in Refs.~\cite{Bultinck2020,Bultinck2020a,Liu2019b, Zhang2020,Abouelkomsan2020}. The resulting single-band-projected interaction Hamiltonian then reads 
\begin{equation}
\label{eq:H_int}
	H_{\text{int}} = \frac{1}{2}\sum_{\substack{\mathbf{k}_1, \mathbf{k}_2, \mathbf{q}}} V(\mathbf{k}_1, \mathbf{k}_2, \mathbf{q}) c_{\mathbf{k}_1}^\dagger c_{\mathbf{k}_2}^\dagger c_{\mathbf{k}_2^{\pd} - \mathbf{q}}c_{\mathbf{k}_1 + \mathbf{q}}^{\pd}\,,
\end{equation}
where $c_{\mathbf{k}}^{\dagger}$ ($c_{\mathbf{k}}^{\pd}$) denotes the creation (annihilation) operator of band electrons in an orbital with momentum $\mathbf{k} \in \text{MBZ}$. The matrix elements are defined as 
\begin{equation}
\label{eq:matrix_elements}
\begin{aligned}
	V(\mathbf{k}_1, \mathbf{k}_2, \mathbf{q}) &= \sum_{\mathbf{G}} V(\mathbf{q} + \mathbf{G}) \Lambda_{\mathbf{k}_1}^{\mathbf{q} + \mathbf{G}} \Lambda_{\mathbf{k}_2}^{-\mathbf{q} - \mathbf{G}}\,,\\
	\text{and} \quad \Lambda_{\mathbf{k}}^{\pm\mathbf{q} \pm \mathbf{G}} &= \left\langle u(\mathbf{k}) | u(\mathbf{k} \pm \mathbf{q} \pm \mathbf{G})\right\rangle
\end{aligned}	
\end{equation}
represents form factors that contain overlaps of the band eigenvectors ${|u(\mathbf{k})\rangle}$ and the summation is over the moir\'{e} reciprocal lattice vectors $\mathbf{G}$ of the continuum model discretization. In the band diagonal basis, the kinetic term takes the simple form ${H_{\text{kin}} = \sum_{\mathbf{k}} \epsilon_{\mathbf{k}} c_{\mathbf{k}}^\dagger c_{\mathbf{k}}^{\pd}}$. Although we start off by assuming a completely flat band and thus neglect $H_{\text{kin}}$, in later sections we account for a finite dispersion by a convex combination of the kinetic and interaction Hamiltonians as $H = \eta H_{\text{kin}} + (1-\eta) H_{\text{int}}$. This is physically equivalent to altering the permittivity $\epsilon$, but leaves the energy scale of the problem approximately constant. This simplifies the interpretation of spectra and provides additional numerical stability. An effective dielectric constant is thus given by ${\epsilon^\ast = \epsilon \eta/(1-\eta)}$ which matches $\epsilon$ at ${\eta=0.5}$. In the case of pure interactions (${\eta = 0}$) we use units of energy ${(e^2/8 \pi \epsilon \epsilon_0 L^\text{M})}$, whereas for the combined Hamiltonian we use \si{meV}. 

Upon performing a particle-hole transformation, the interaction Hamiltonian induces a single-hole dispersion \cite{Abouelkomsan2020,Laeuchli2013}
\begin{align}
\label{eq:idh}
	E_h(\mathbf{k}) = \frac{1}{2}\sum_{\mathbf{k}^\prime} V_{\mathbf{k}^\prime \mathbf{k} \mathbf{k}^\prime \mathbf{k}} + V_{\mathbf{k} \mathbf{k}^\prime \mathbf{k} \mathbf{k}^\prime} - V_{\mathbf{k} \mathbf{k}^\prime \mathbf{k}^\prime \mathbf{k}}- V_{\mathbf{k}^\prime \mathbf{k} \mathbf{k} \mathbf{k}^\prime}\,,\nonumber\\
	V_{\mathbf{k}_1 \mathbf{k}_2 \mathbf{k}_3 \mathbf{k}_4} = V_{\mathbf{k}_1 \mathbf{k}_2 \mathbf{k}_2 - \mathbf{q} \mathbf{k}_1 + \mathbf{q}} \equiv V(\mathbf{k}_1, \mathbf{k}_2, \mathbf{q})\,,\quad\quad
\end{align}	
which turns out to be a useful characteristic for the intuitive understanding of one important aspect of the Coulomb interaction structure in this system. This (sign flipped) induced hole dispersion (IHD) is evaluated in the bottom row of \cfig{fig:overview}. As $w_{AB}$ varies, the features of $-E_h(\mathbf{k})$ remain qualitatively similar with a pronounced maximum at $\mathbf{\Gamma}$. The same holds for the situation in the conduction band. The IHD of ${w_{AB}=\SI{110}{meV}}$ differs from the one with ${w_{AB}=\SI{90}{meV}}$ primarily by a larger bandwidth.

Figure~\ref{fig:overview} suggests that the vital differences in the model are captured in the cases ${w_{AB}=\SI{90}{meV}}$ and ${w_{AB}=\SI{110}{meV}}$, while intermediate values smoothly interpolate between these scenarios. We thus restrict ourselves to the two representative cases ${w_{AB}=\SI{90}{meV}}$ and $\SI{110}{meV}$ in our exact diagonalization study of the many-body Hamiltonian.

\section{Numerical method and signatures of correlated phases \label{sec:method}} 

Similarly to Refs.~\cite{Abouelkomsan2020, Repellin2020, Xie2020a}, we use Lanczos based exact diagonalization (ED) in momentum space to tackle the many-body problem of interacting band-fermions. This enables us to obtain numerically \emph{exact} ground-state energies as well as measurements of observables on finite size clusters with various geometric features at arbitrary filling fractions. In the considered spin- and valley-polarized subsector of a single band, the total Hilbert space dimension for a given number of electrons $N_e$ on $N_k$ orbitals is the binomial coefficient~$\binom{N_k}{N_e}$. By utilizing the translational symmetry of the system, the total Hilbert space decomposes into subspaces of $N_k$ center of mass (COM) momenta $\mathbf{k}_{\text{COM}} = \sum_{i=1}^{N_e} \mathbf{k}_i$. In order to keep the code applicable to general geometries and Hamiltonians of potentially reduced symmetry, no point group operations are exploited in the algorithm. The average linear matrix dimension is then $\binom{N_k}{N_e}/N_k$, which culminates in about 252 million states in the study of the cluster \nameref{cl:6.0.0.6} at ${\nu=1/2}$.

The algorithm provides access to the ground-state wave function(s) as well as the momentum orbital resolved low-energy spectrum. This is a key advantage of the ED method, as many phases have distinct signatures in the structure of the low-energy spectrum, e.g., in the k-space location and degree of quasi-degenerate ground-state energy levels. To be precise, it should be noted that exact degeneracy generally holds only in the thermodynamic limit (TDL) and a finite splitting of the ground-state manifold is to be expected on finite clusters. With regard to the phases encountered in the current study, at a filling of $\nu=p/q$ the FCI manifests in the spectrum via a $q$-fold degeneracy of orbitals satisfying the generalized Pauli principle developed in \cref{Regnault2011}, which was extended to a heuristics on more general clusters in \cref{Laeuchli2013}. In addition, the ground-state orbitals of an FCI are expected to exhibit spectral flow, that is, under the insertion of magnetic flux quanta like $\mathbf{k} \rightarrow \mathbf{k} + \frac{\Phi_i}{2 \pi} \mathbf{g}_{\mathbf{k},i}$ they exchange their order without mixing with excited states. The flow of orbitals into each other may be hindered if all $q$ ground-states coincide in their COM momentum. Nevertheless, they should remain isolated from higher lying states and the original spectrum has to be restored at $\Phi_i/2 \pi=q$. On the other hand, the degeneracy of the CDW depends on the specific pattern that is realized, i.e. in what manner the spatial symmetry is broken. For a simple WC-like order (with a single orientation) at $\nu=1/q$, the ground-state is $q$-fold degenerate, with orbitals separated by the order vectors $\{\mathbf{q}^{\ast}_i\}$. The possible variants of more complex patterns need to be counted individually, however the spontaneous symmetry breaking aspect may generally be analyzed using group theoretical tools in order to predict the location of ground-state COM orbitals \cite{Wietek2017}. A particular pattern manifests in the spatial correlations of the charge density, which can be measured using the charge structure factor. In the considered single-band-projected setting, we define it as
\begin{equation}
	\label{eq:Sq}
	\begin{aligned}
		S(\mathbf{q}) \equiv \frac{1}{N_k} &\left[\sum_{\mathbf{k}} \left|\Lambda_{\mathbf{k}}^{\mathbf{q}}\right|^2 n(\mathbf{k}) +  \right. \\
		&\;\left. \sum_{\mathbf{k}_1,\mathbf{k}_2} \Lambda_{\mathbf{k}_1}^{\mathbf{q}} \Lambda_{\mathbf{k}_2}^{-\mathbf{q}}\left\langle c^\dagger_{\mathbf{k}_1} c^\dagger_{\mathbf{k}_2} c_{\mathbf{k}_2^{\pd} - \mathbf{q}} c_{\mathbf{k}_1 + \mathbf{q}}^{\pd} \right\rangle \right],
	\end{aligned}	
\end{equation}
where $n(\mathbf{k}) \equiv \langle c^\dagger_{\mathbf{k}} c^{\pd}_{\mathbf{k}} \rangle$ is the orbital occupation. A detailed derivation may be found in \capp{sec:derivation_Sq}.

\section{Competition of correlated phases at canonical filling fractions \label{sec:nu_canonical}}

We start our discussion of numerical results with the single-band model at ${\nu=1/3}$ filling. We find strong evidence for the manifestation of an FCI for the band parameter ${w_{AB}=\SI{90}{meV}}$, therefore corroborating the findings in Ref.~\cite{Abouelkomsan2020}, and a translation symmetry breaking CDW with order wave vector $\mathbf{q}^\ast=\mathbf{K}_{\pm}$ at ${w_{AB}=\SI{110}{meV}}$, which is at odds with Ref.~\cite{Abouelkomsan2020}. We provide an explanation for the underlying order mechanisms and investigate the stability of the phases against the introduction of kinetic energy. After highlighting the differences and commonalities in the conduction band and for the complementary filling ${\nu=2/3}$, we turn to the investigation of other potentially interesting fractions. We reveal a series of WCs at commensurate fillings $\nu<1/3$ as well as the realization of a second FCI at ${\nu=2/5}$, confirming similar calculations in \cref{Repellin2020}. Finally, we present our numerical results for half filling of the moir\'{e} flat band, which, however, do not allow us to conclusively determine the nature of the ground-state.

\subsection{FCI versus CDW at ${\nu=1/3}$  \label{sec:fci_cdw}}

\begin{figure*}[htp]
	\includegraphics[width=0.9\textwidth]{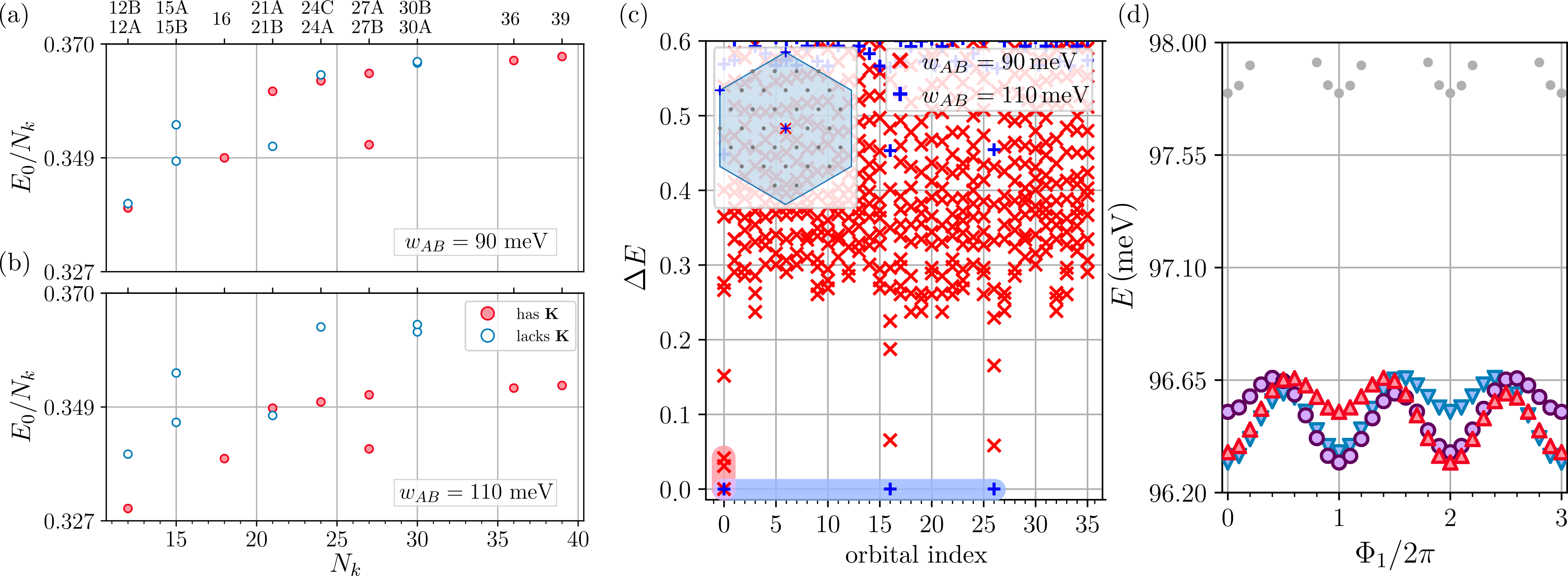}
	\caption{(a,b) Ground-state energy for various clusters and (c) many-body spectrum on the cluster \nameref{cl:6.0.0.6} at $\nu= 1/3$ filling for both band parameters. At ${w_{AB}=\SI{110}{meV}}$ a clear sensitivity to the presence of the $\mathbf{K}$ points is observable while the ${w_{AB}=\SI{90}{meV}}$ ground-state in (a) is indifferent to this geometric feature. Clusters with aspect ratios far from 1, like \nameref{cl:3.0.0.7} and \nameref{cl:3.0.0.9}, violate this pattern. For more geometric details see \ctab{tab:clusters}. Shaded areas in (c) mark the set of identified ground-states, whose locations in the MBZ are marked in the inset for the respective hopping amplitude. The k-space locations of momenta associated to each orbital index are displayed in \capp{sec:spectra}. (d) Spectral flow of ground-state orbitals on the cluster \nameref{cl:1.4.5.-1} at ${w_{AB}=\SI{90}{meV}}$ under the insertion of magnetic flux $\Phi_1$. We incorporated a slight valence band dispersion via $\eta=0.3$ for better separation from excited states. The general effect of kinetic energy is discussed in \csec{sec:interplay}.
		\label{fig:observables_nu_1_over_3}}
\end{figure*}

To begin with, we consider the pure interaction Hamiltonian of \ceqn{eq:H_int} and compute the low-lying eigenvalues and eigenvectors on various cluster geometries detailed in \capp{sec:clusters}. Figure~\ref{fig:observables_nu_1_over_3}(a,b) displays the obtained ground-state energies per orbital over the system size for both considered interlayer hopping amplitudes. While \cfig{fig:observables_nu_1_over_3}(a) is fairly featureless up to a gradual convergence of the ground-state energy with increasing system size, \cfig{fig:observables_nu_1_over_3}(b) signals a pronounced sensitivity to the presence of the $\mathbf{K}_{\pm}$ points, with the ground-state energy being lower when the $\mathbf{K}_{\pm}$ points are present. This points to a different phase than an FCI, whose ground-state energy is expected to be rather insensitive to the global cluster shape (within reasonable limits).

The obtained many-body spectra, such as \cfig{fig:observables_nu_1_over_3}(c), reveal an approximate three-fold degenerate ground-state, where the COM momentum orbitals in the
ground-state manifold are found to be distinct among the two considered interlayer hopping amplitudes on multiple clusters.
\begin{figure}[hbp]
	\includegraphics[width=0.95\columnwidth]{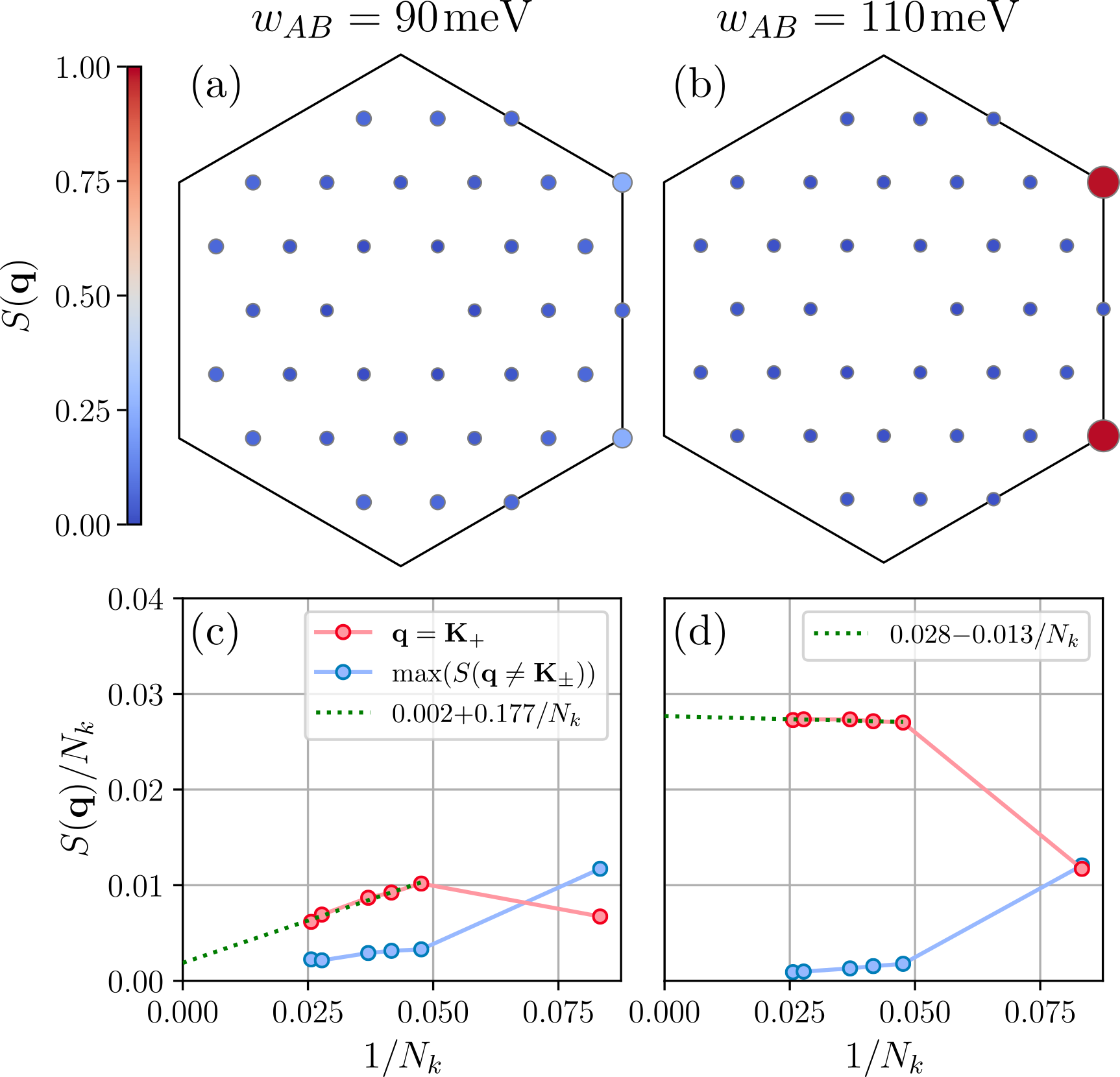}
	\caption{Structure factor distribution over the MBZ of the cluster \nameref{cl:6.0.0.6} and extrapolation to the TDL for $\nu= 1/3$ and both $w_{AB}$. The dominant peaks in (b) are strong evidence for $\mathbf{K}$-CDW order and the accompanied tripling of the unit cell, which survives in the TDL in (d). The CDW signatures at ${w_{AB}=\SI{90}{meV}}$ in (a) and (c) are less pronounced and are expected to vanish in the TDL. 
		\label{fig:Sq_nu_1_over_3}}
\end{figure}
While at ${w_{AB}=\SI{90}{meV}}$ they follow the ${\nu=1/3}$ FCI heuristics \cite{Regnault2011,Laeuchli2013}, the ground-state momenta at ${w_{AB}=\SI{110}{meV}}$ are separated by the moir\'{e} Dirac point momenta $\mathbf{K}_{\pm}$ (on clusters which feature these points in the MBZ). Upon inserting $\Phi_1$ flux quanta, in \cfig{fig:observables_nu_1_over_3}(d) we observe that the three ground-states at ${w_{AB}=\SI{90}{meV}}$ flow into each other without mixing with higher excited states. At ${\Phi_1/2\pi = 3}$, the original spectrum is restored, in accordance with Laughlin-like states at filling ${\nu=1/3}$. It should be noted that we intentionally chose a cluster with three distinct ground-state momenta in the FCI phase to enable proper spectral flow, which also features the $\mathbf{K}_{\pm}$ points. 

The spectral analysis and energetic considerations point towards the possibility of different types of order depending on the interlayer hopping amplitude $w_{AB}$. Where the data at ${w_{AB}=\SI{90}{meV}}$ suggests the formation of a topological fractional Chern insulator, in accordance with the results of \cref{Abouelkomsan2020} and \cref{Repellin2020}, ${w_{AB}=\SI{110}{meV}}$ appears to favor order whose signatures are consistent with charge-density-waves with order momentum $\mathbf{K}_{\pm}$. 
\begin{figure}[bp]
	\begin{minipage}{\linewidth}
		\begin{minipage}[b]{.47\columnwidth}
			\includegraphics[height=0.14\paperheight]{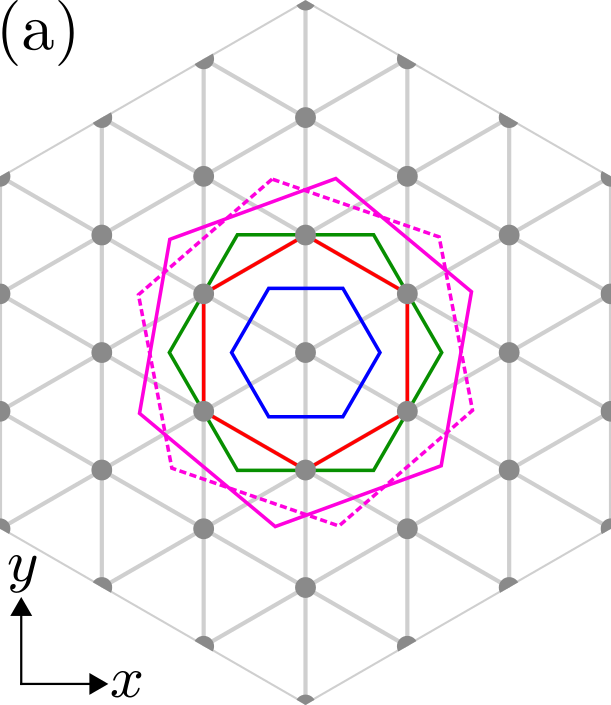}
		\end{minipage}
		\begin{minipage}[b]{.47\columnwidth}
			\includegraphics[height=0.14\paperheight]{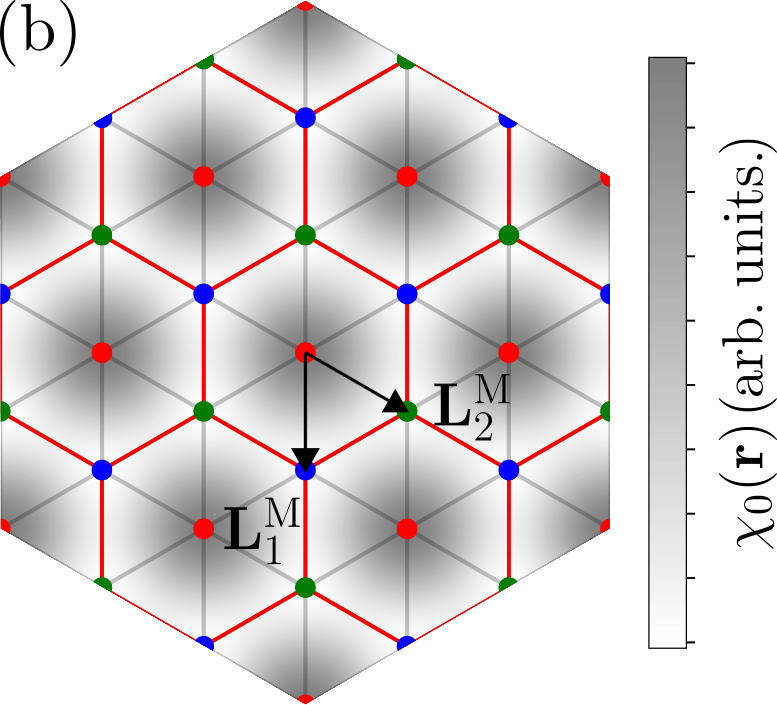}
		\end{minipage}
	\end{minipage}	
	\caption{(a) Various commensurate enlargements of the original moir\'{e} Wigner-Seitz cell (blue), corresponding to CDW/WCs that break the real-space $\mathbf{L}_{1,2}^\text{M}$ moir\'{e} translational symmetry. The respective fillings are ${\nu=1/3}$ (red), ${\nu=1/4}$ (green) and ${\nu=1/7}$ (magenta). The pattern at ${\nu=1/7}$ splits into two classes, which are related by an out-of-plane $C_2$ operation. (b) Illustration of the characteristic density-density correlation function $\chi_0(\mathbf{r})$ as well as the three orthogonal realizations (red, green, blue) of a CDW at ${\nu=1/3}$ (cf. \capp{sec:derivation_Sq} for details).
		\label{fig:cdw_illustration}}
\end{figure}
The emergence of CDW order is reflected most prominently in the charge structure factor of \ceqn{eq:Sq} in \cfig{fig:Sq_nu_1_over_3}(a,b): Little spatial modulation is present for $\SI{90}{meV}$, where for ${w_{AB}=\SI{110}{meV}}$ the hallmark Bragg peaks of a CDW manifest at momenta $\mathbf{K}_{\pm}$. The finite size extrapolation of the peak height to the TDL in \cfig{fig:Sq_nu_1_over_3}(c,d) shows that long-range order is stable, while the signal off the order momentum vanishes. Although the order parameter in \cfig{fig:Sq_nu_1_over_3}(c) also extrapolates to a nominally finite value for ${w_{AB}=\SI{90}{meV}}$, it is significantly smaller than in \cfig{fig:Sq_nu_1_over_3}(d) and will most likely approach zero for larger clusters, in accordance with the expectations for an FCI state.

The $\mathbf{K}$-CDW can be imagined in real-space as illustrated in \cfig{fig:cdw_illustration}. It is the first in a series of Wigner crystal-like states that are locked to the underlying moir\'{e} triangular lattice and spontaneously break translational symmetry, thus leading to an enlargement of the moir\'{e} unit cell. At filling ${\nu=1/3}$, the unit cell is tripled, such that each of the three degenerate ground-states corresponds to one realization on the triangular moir\'{e} lattice of \cfig{fig:cdw_illustration}(b).

\subsection{Interplay of Berry curvature, induced hole dispersion and kinetic energy \label{sec:interplay}}

\begin{figure}[b]
	\includegraphics[width=0.9\columnwidth]{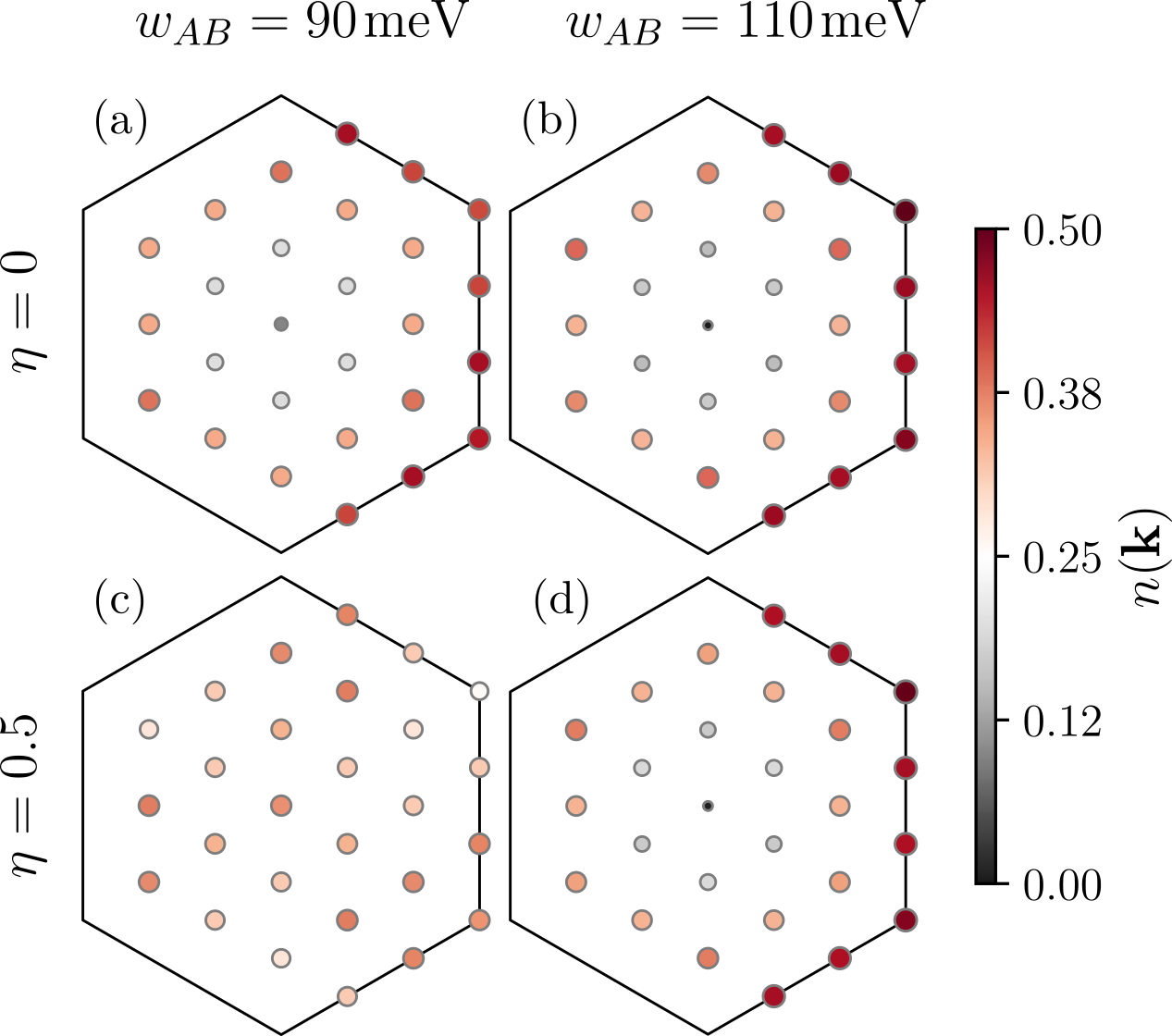}
	\caption{Orbital occupation at $\nu= 1/3$ for both $w_{AB}$ and two convex combination factors $\eta$. While (a) and (c) indicate a smoothed redistribution of $n(\mathbf{k})$ with $\eta$, (b) and (d) are nearly identical. The used cluster is \nameref{cl:3.3.3.-6}. 
		\label{fig:occupation_nu_1_over_3}}
\end{figure}

What differentiates the situation at ${w_{AB}=\SI{90}{meV}}$ from the one at ${w_{AB}=\SI{110}{meV}}$, such that either the formation of the FCI or the CDW is favored? We can gain insight into the driving mechanism by studying the orbital occupation $n(\mathbf{k})$, which informs us about the predominant locations of the electrons in the MBZ. In the pure interaction case of \cfig{fig:occupation_nu_1_over_3}(a) and \cfig{fig:occupation_nu_1_over_3}(b), it tells us that the Coulomb repulsion depletes the region near $\mathbf{\Gamma}$ and redistributes the electrons towards the border of the MBZ, an effect previously discussed in Refs.~\cite{Laeuchli2013,Abouelkomsan2020}. 
This can be intuitively understood in the hole picture, where the IHD in the lowest panels of \cfig{fig:overview} encourages holes being close to $\mathbf{\Gamma}$. At a hole filling fraction of ${\nu_h=2/3}$, most of the inner region of the MBZ is occupied by holes, while electrons are closer to the boundary. The increased IHD bandwidth for ${w_{AB}=\SI{110}{meV}}$ leads to an amplified interaction driven reallocation of electrons to the outer orbitals when compared to ${w_{AB}=\SI{90}{meV}}$.

\begin{figure}
	\includegraphics[width=0.95\columnwidth]{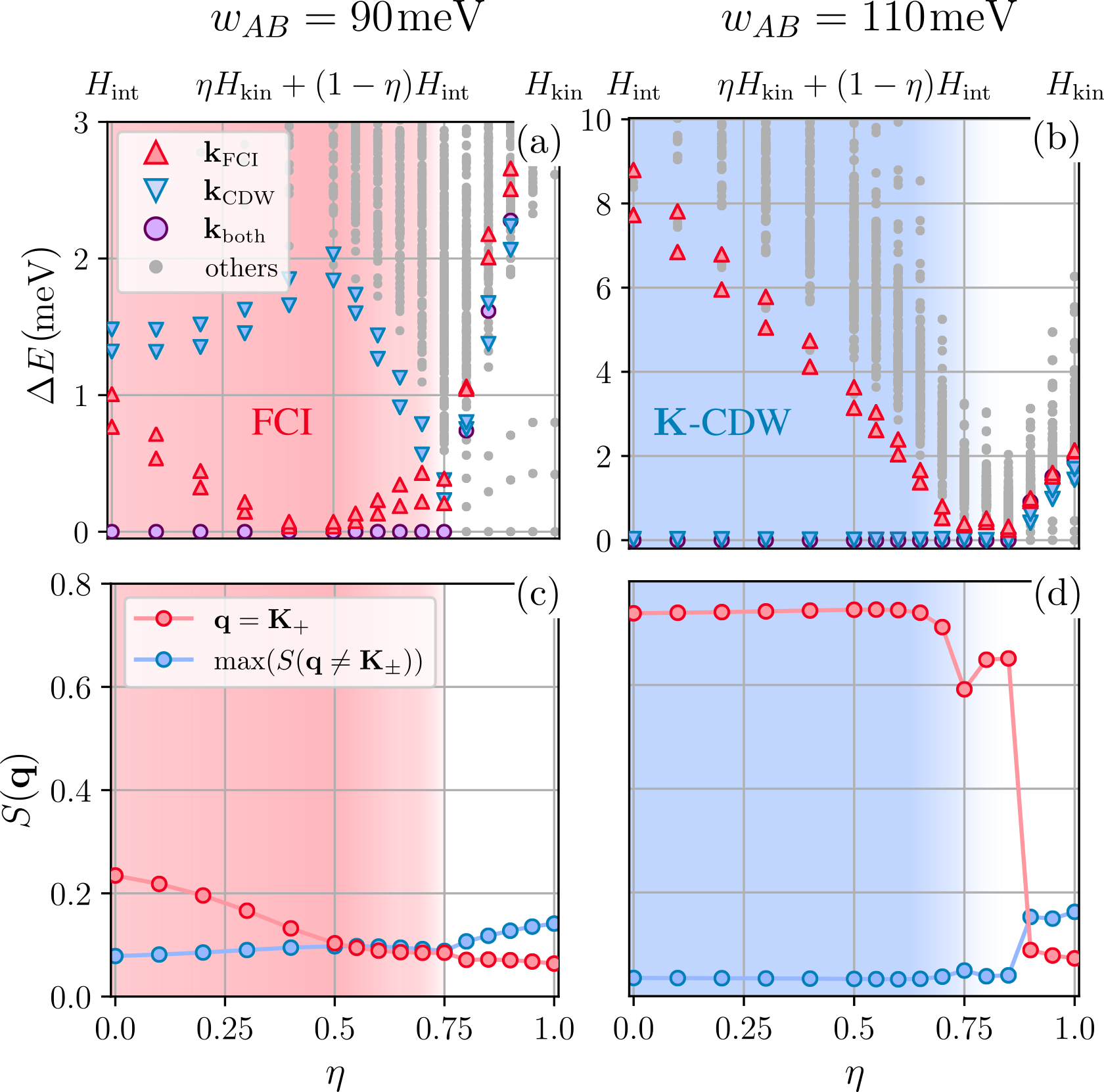}
	\caption{(a) The FCI reaches a stability maximum near $\eta\simeq0.5$, accompanied by a suppression of $S(\mathbf{q}=\mathbf{K}_{\pm})$. The stable CDW is degraded with $\eta$ until its signatures vanish close to $\eta\simeq0.8-0.9$. The used cluster is \nameref{cl:3.3.3.-6}.
	\label{fig:convex_spectrum_Sq}}
\end{figure}

The crucial difference between the two cases is however that for ${w_{AB}=\SI{110}{meV}}$, most of the Berry curvature in \cfig{fig:overview}(e) is concentrated close to $\mathbf{\Gamma}$, while the electrons arrange at the border of the MBZ. Thus they do not experience a significant effective magnetic field, which would otherwise encourage the formation of a FQH-like state, and charge order by a tripling of the unit cell is the energetically most favourable option, with the appealing real-space interpretation of minimizing the Coulomb interactions by maximizing the distance between the electrons. The large gap in the spectrum to the COM orbitals dictated by the FCI heuristics in \cfig{fig:observables_nu_1_over_3}(c) as well as \cfig{fig:convex_spectrum_Sq}(b) affirm the robustness of the $\mathbf{K}$-CDW. In contrast to the authors of \cref{Abouelkomsan2020} who proposed an FCI for ${w_{AB}=\SI{110}{meV}}$ at a reduced screening length of ${\lambda=L^{\text{M}}/6}$, we observe, for the same parameter set, clear signatures of CDW order in spectra such as \cfig{fig:spectrum_ktot_nu_1_over_3_LM6th} in \capp{sec:spectra}, as well as the structure factor on various clusters. Although the pure interaction orbital occupation is practically the same as for ${w_{AB}=\SI{90}{meV}}$, the Berry curvature in the latter case is distributed more uniformly as shown in \cfig{fig:overview}(a). 

The rather small excitation energies to COM orbitals corresponding to the CDW and the poor degeneracy of FCI ground-states in \cfig{fig:observables_nu_1_over_3}(c) and \cfig{fig:convex_spectrum_Sq}(a) suggest a close competition between these phases on lattices that geometrically support the $\mathbf{K}$-CDW. 

We have established that for a completely flat band an FCI is the most likely ground-state for ${w_{AB}=\SI{90}{meV}}$, while the ${w_{AB}=\SI{110}{meV}}$ configuration favors CDW order. The effect of a finite kinetic energy bandwidth is now to be discussed by including the continuum model valence band dispersion via $H_{\text{kin}}$. As described in \csec{sec:model}, this is done by a convex combination of $H_{\text{kin}}$ and $H_{\text{int}}$ controlled by the parameter ${\eta \in \left[ 0,1 \right]}$. The pure interaction case is obtained for ${\eta=0}$, while ${\eta=1}$ leads to a non-interacting Hamiltonian containing only the kinetic energy. In \cfig{fig:convex_spectrum_Sq} we observe the behavior of both the spectrum and the CDW order parameter while varying $\eta$ from $0$ to $1$. The inspection of \cfig{fig:convex_spectrum_Sq}(b) and \cfig{fig:convex_spectrum_Sq}(d) suggests that the long-range CDW order is gradually penalized by the kinetic energy until the spectral gap and the dominance of ${S(\mathbf{q}=\mathbf{K}_+)}$ vanish around ${\eta=0.8\mbox{--}0.9}$. This corresponds to an effective relative permittivity of ${\epsilon^\ast = 2.675 \times \eta /(1-\eta) \simeq 10\mbox{--}24}$, which is above experimental estimates for bilayer graphene interfaces at ${\epsilon^{\ast} = 6 \pm 2}$ (${\eta \simeq 0.6\mbox{--}0.75}$)~\cite{Bessler2019}. A picture that might seem peculiar at first glance emerges from \cfig{fig:convex_spectrum_Sq}(a). Here the single-particle dispersion does not appear to immediately weaken the FCI, but the ratio of the excitation gap to the ground-state splitting improves until ${\eta = 0.4\mbox{--}0.5}$. This is at odds with the canonical view that a general interaction driven phase profits from a band that is as flat as possible. The origin of this curious feature may be understood in terms of the role of the single-particle dispersion in the previously developed mechanism for the manifestation of FCI or CDW states. The crucial aspect of the valence band is the minimum at $\mathbf{\Gamma}$ while the outer momenta remain higher in energy. In a non-interacting setting, the lowest orbitals in the trough would be successively filled to accommodate all $N_e$ electrons and consequently form a Fermi surface. Turning on the Coulomb repulsion, the IHD shifts the electron density towards the border of the MBZ and thus, in the valence band, acts opposite to the preference of the single-particle dispersion. Such a scenario generates a sweet spot, where electrons are almost uniformly distributed across the MBZ. What is more, the electrons in the FCI state can now take advantage of the effective magnetic field, that is the Berry curvature of \cfig{fig:overview}(a), across the whole MBZ.

Increasing $\eta$ past $0.5$ degrades the FCI and leads to an almost degenerate situation of FCI and CDW momentum orbitals near ${\eta=0.75}$, from which onward the orbital occupation distribution appears to be dictated by the band structure alone. In principle, such a sweet spot may also be present in \cfig{fig:convex_spectrum_Sq}(b), yet upon closer inspection of the situation at ${\eta \simeq 0.8}$, no clear signature of an FCI was observed. The reason might be that the optimal $\eta$ is quite far below the value at which $H_{\text{kin}}$ perturbs the CDW enough for the FCI to compete. Interestingly for the CDW, as indicated by \cfig{fig:occupation_nu_1_over_3}(d) and \cfig{fig:convex_spectrum_Sq}(d), $n(\mathbf{k})$ and also $S(\mathbf{q}=\mathbf{K}_+)$ are practically unaffected by an increase of $\eta$ until the start of the breakdown of CDW order at $\eta=0.75$. This suggests a high degree of stability of the $\mathbf{K}$-CDW wave function across a large interval of kinetic energy strengths.

\subsection{Conduction band physics and complementary filling\label{sec:complementary}}

\begin{figure}[b]
	\includegraphics[width=0.95\columnwidth]{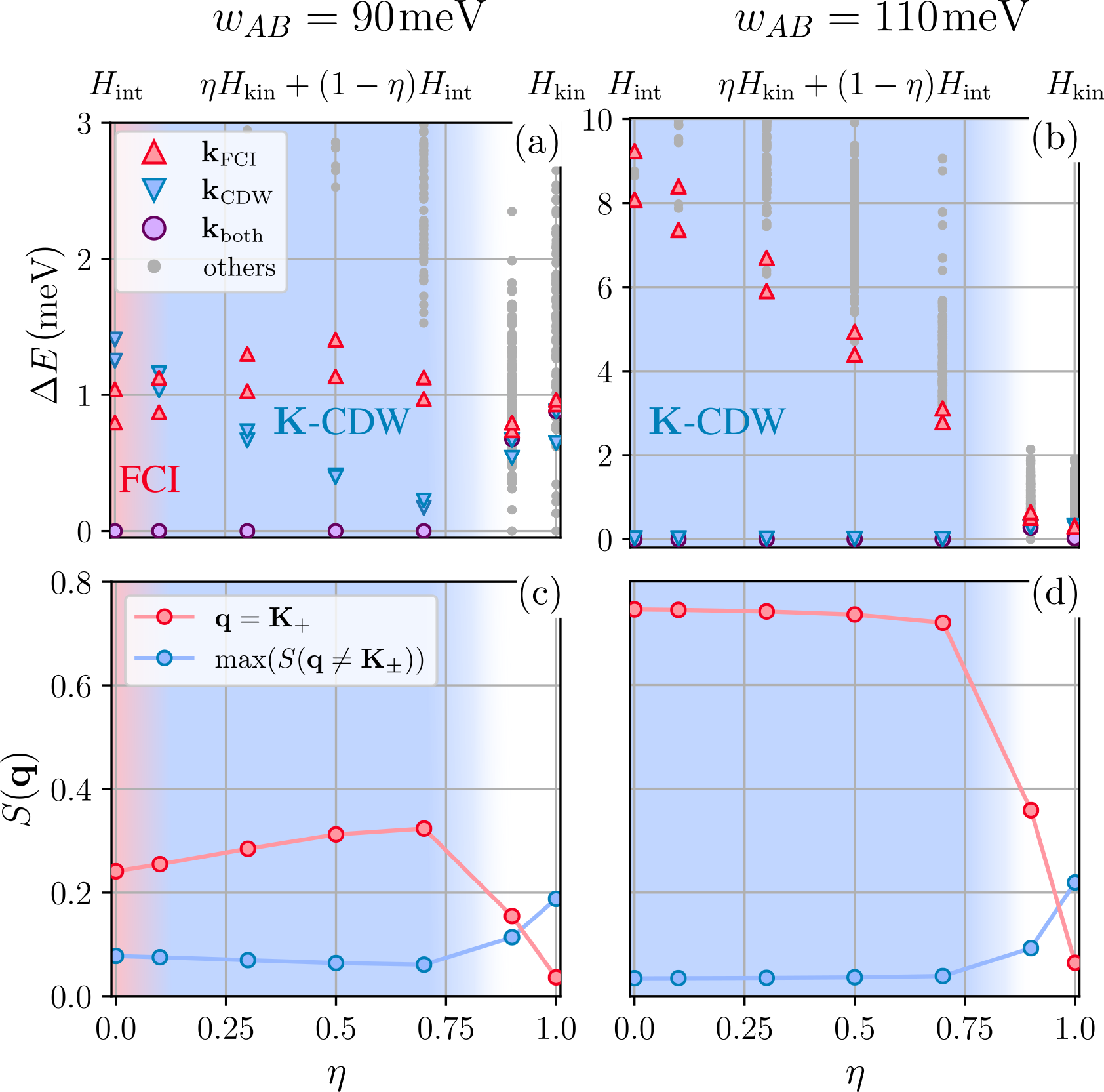}
	\caption{(a) Increasing $\eta$ in the conduction band at $\nu=1+1/3$ leads to a suppression of the FCI for ${w_{AB}=\SI{90}{meV}}$, stabilizing the CDW according to (c) up until $\eta\simeq0.7$. The data in (b) and (d) at ${w_{AB}=\SI{110}{meV}}$ qualitatively replicate the situation of \cfig{fig:convex_spectrum_Sq}(b,d), albeit the CDW order is slightly more stable. The used cluster is \nameref{cl:3.3.3.-6}. 
		\label{fig:convex_spectrum_Sq_conduction}}
\end{figure}

We now investigate the similarities and differences when switching to the conduction band (${\nu=1+1/3}$) and upon adding twice the amount of electrons to a single (valence or conduction) band (${\nu=2/3}$). The effective reflection about zero energy results in the observed energetic peak of the conduction band at $\mathbf{\Gamma}$ in \cfig{fig:overview}, which falls off towards the MBZ border. The Berry curvature in the conduction band of the same valley is related by a sign and coordinate flip to the one in the valence band (up to a slight particle-hole asymmetry). Thus the ones displayed in \cfig{fig:overview} properly represent the magnitude at the center and the border of the MBZ, which suffices for our discussion. The IHD is almost identical up to a reflection about a $\mathbf{\Gamma}$-$\mathbf{K}$ path and thus has the same qualitative effect as in the valence band. In our numerical results, we first compare the data for ${\eta=0}$ in \cfig{fig:convex_spectrum_Sq_conduction} to the same set of points in \cfig{fig:convex_spectrum_Sq}. 
\begin{figure*}[t]
	\centering
	\includegraphics[width=0.9\textwidth]{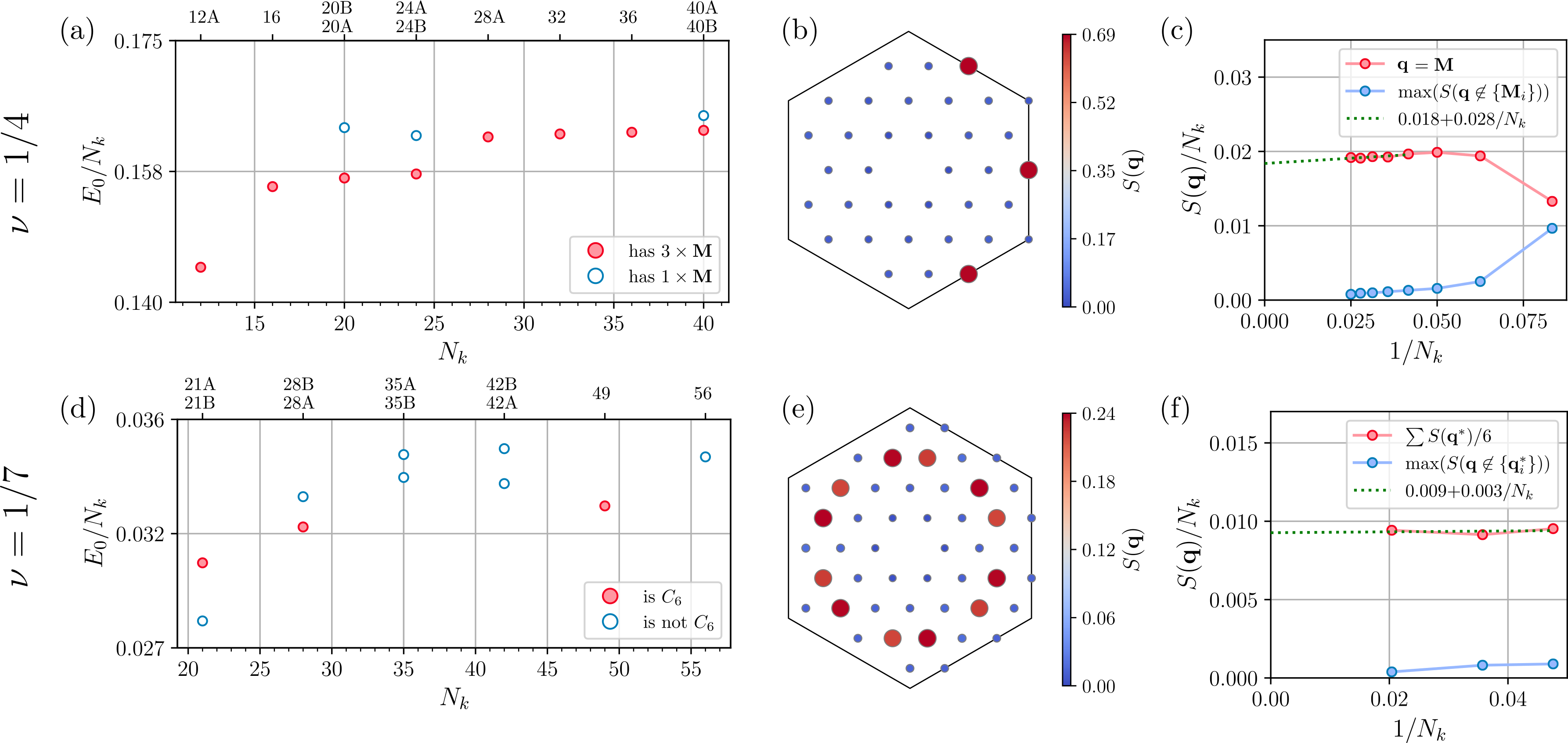}
	\caption{Evidence for an $\mathbf{M}$-WC at ${\nu=1/4}$ (a)--(c) and a $C_6$-WC at ${\nu=1/7}$ (d)--(f) for ${w_{AB}=\SI{110}{meV}}$. Although the signatures are less pronounced than for ${\nu=1/3}$, they are clearly visible in (b) and (e). The finite-size data in (c) and (f) suggest the order will prevail in the TDL. The displayed discretizations in (b) and (e) are \nameref{cl:6.0.0.6} and \nameref{cl:7.0.0.7} respectively. For more details on the used clusters, refer to \ctab{tab:clusters}.
		\label{fig:observables_nu_smaller}}
\end{figure*}
The results (FCI at ${w_{AB}=\SI{90}{meV}}$, CDW at ${w_{AB}=\SI{110}{meV}}$) almost exactly coincide, which is the consequence of the time-reversal and particle-hole relations for bands from different valleys. More remarkable behavior that distinguishes the two bands is revealed when tuning ${\eta>0}$. While for ${w_{AB}=\SI{110}{meV}}$ we arrive at results that are reminiscent of \cfig{fig:convex_spectrum_Sq}(b) and \cfig{fig:convex_spectrum_Sq}(d), depicting an even slightly more stable $\mathbf{K}$-CDW that is slowly disfavored by the kinetic energy, \cfig{fig:convex_spectrum_Sq_conduction}(a) and \cfig{fig:convex_spectrum_Sq_conduction}(c) show no signs of a further stabilization of the FCI. On the contrary, both, the spectra and the order parameter $S(\mathbf{q})$ signal that the CDW profits from increasing $\eta$ until about 0.7. 
Therefore, although the ground-state for pure interactions appears to be an FCI, it is quickly suppressed near ${\eta=0.1}$, which corresponds to ${\epsilon^\ast \simeq 0.3}$, and the CDW stabilizes throughout an interval ${\eta \in \left(0.1,0.7\right]}$. Intuition is gained by realizing that the only crucial qualitative modification to the valence band situation is an essentially flipped single-particle dispersion, which favors electrons at the MBZ boundary instead of the center. It thus reinforces the effect of the IHD on the orbital occupation and no FCI sweet spot can occur as the electrons are driven away from the Berry curvature at $\mathbf{\Gamma}$ more vigorously.

Finally, we elaborate on the ${\nu=2/3}$ (${1+2/3}$) filling of the valence (conduction) band in the developed framework of the interplay between kinetic energy, Berry curvature and induced hole dispersion. In order to keep this manuscript condensed, we do not display separate results for these configurations. We find that the situation is qualitatively very similar to the filling of ${\nu=1/3}$, albeit the FCI in the valence band at ${w_{AB}=\SI{90}{meV}}$ features a larger excitation gap at ${\eta=0}$, it is again stabilized by the kinetic energy compared to the CDW. The conduction band results suggest the onset of a transition from the FCI towards the CDW order upon increasing the strength of the kinetic energy. However no clear separation as in \cfig{fig:convex_spectrum_Sq_conduction}(a) is present in the limited data for this configuration. Up to a reduction of the robustness, we find clear evidence for CDW order in both bands at ${w_{AB}=\SI{110}{meV}}$. An increase of $\eta$ again gradually closes the excitation gap until it vanishes near ${\eta=0.7}$. Nearly all of the observed features in the ${\nu=2/3}$ data are explainable akin to the situation at $1/3$ (${1+1/3}$) filling. Where for low $\eta$, electrons were almost exclusively located at the border of the MBZ, by the fermionic exclusion principle now twice as much weight has to be accommodated. This leads to an initially more stable FCI and a weakened CDW because more of the overall Berry curvature is experienced by the collective electron wave function. Consequently, the CDW order is destroyed faster but the general dependence on $\eta$ is smoothed because a lower fraction of the total weight of the wave function can be redistributed into a specific region of the MBZ. 

\subsection{Evidence for charge order at lower filling\label{sec:nu_smaller}}

\begin{figure*}[t]
	\centering
	\includegraphics[width=0.9\textwidth]{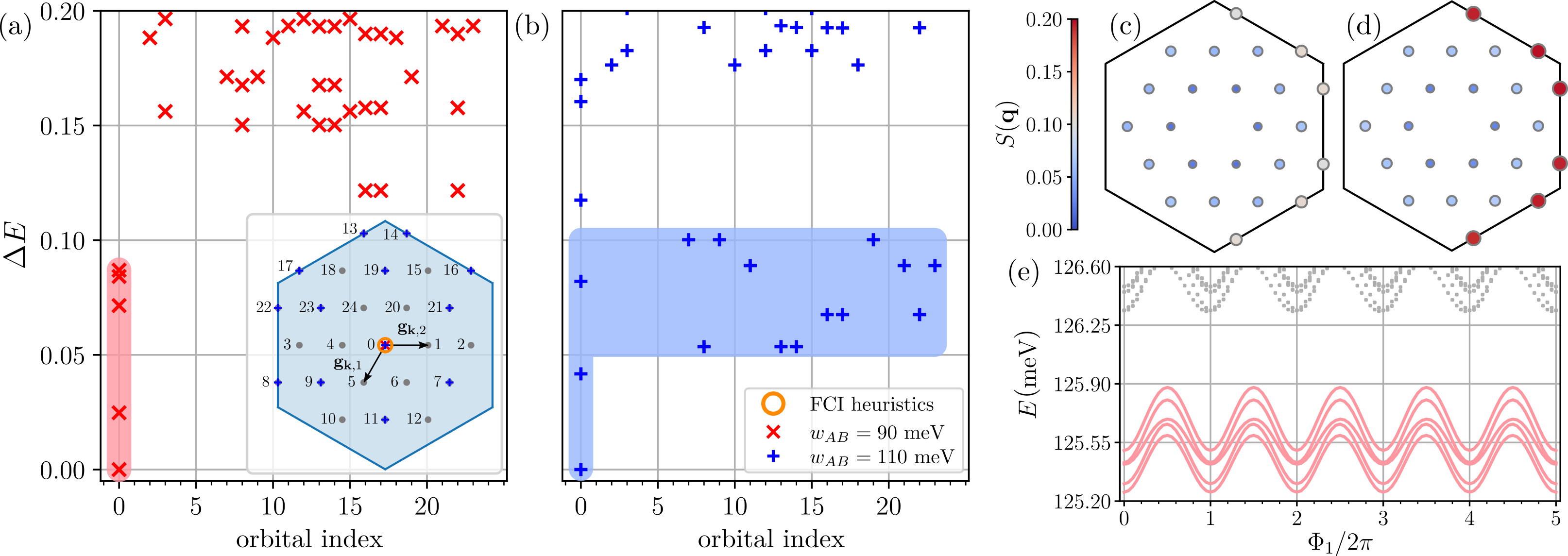}
	\caption{Many-body spectrum at ${\nu=2/5}$ for (a) ${w_{AB}=\SI{90}{meV}}$ and (b) ${w_{AB}=\SI{110}{meV}}$. (c,d) The charge structure factor for both band parameters at ${\nu=2/5}$ as well as (e) spectral flow of the ${w_{AB}=\SI{90}{meV}}$ ground-states, consistent with an FCI. In (e) ${\eta=0.5}$ is used for clearer separation of the ground-state manifold. Symbols have been omitted for clarity. The used cluster is \nameref{cl:5.0.0.5}.  \label{fig:observables_nu_2_over_5}}
\end{figure*}
Motivated by our findings of robust CDW order at ${\nu=1/3}$, we analyze the possibility of states with even larger interaction induced unit cells. Fig.~\ref{fig:cdw_illustration} visualizes the next larger four- and sevenfold extensions of the moir\'{e} unit cell, corresponding to band fillings of ${\nu=1/4}$ and ${\nu=1/7}$ respectively. The pattern at ${\nu=1/4}$ translates exactly along the doubled moir\'{e} lattice vectors, which would imply a charge order vector of $\mathbf{q}^\ast=\mathbf{M}_i$, where the index $i$ denotes the possibility of three inequivalent $\mathbf{M}$ points in the MBZ. We thus dub this order the $\mathbf{M}$-WC. The WC at ${\nu=1/7}$ extends even further, such that seven individual moir\'{e} sites are contained within the WC unit cell of \cfig{fig:cdw_illustration}. A peculiarity here are the two possible, inequivalent realizations of this spatial modulation on the triangular lattice, which are related by an out of plane $C_2$ operation along a moir\'{e} lattice vector. Because the real-space translation vector of the order is even larger in magnitude than for ${\nu=1/3}$ or ${\nu=1/4}$, the corresponding order momenta have to be located inside the MBZ. We would expect the charge order parameter $S(\mathbf{q})$ to develop substantial peaks at six momenta $\mathbf{q}^\ast$ for each realization of the WC pattern. On clusters with $C_6$ symmetry, only one pattern may be realized while a $D_6$ symmetric cluster supports superpositions of both WC orientations, which makes a total of 12 potential order momenta and a 14-fold ground-state degeneracy instead of the expected sevenfold. The order momenta and ground-state orbitals then fall into two classes, where within each the nonzero orbital and order momenta are related by a $C_6$ operation. This type of WC is henceforth referred to as the $C_6$-WC. Because we arrive at qualitatively the same results for both considered values of $w_{AB}$, we discuss only the more pronounced situation at ${w_{AB}=\SI{110}{meV}}$. In \capp{sec:spectra}, the many-body spectra for both hopping amplitudes and fillings are displayed in \cfig{fig:spectrum_ktot_nu_1_over_4},  \cfig{fig:spectrum_ktot_nu_1_over_7}(a) and \cfig{fig:spectrum_ktot_nu_1_over_7}(b).

We now take a look at the data presented in \cfig{fig:observables_nu_smaller}. Starting with the filling ${\nu=1/4}$, a slight energetic advantage appears to be present in \cfig{fig:observables_nu_smaller}(a) for clusters that realize all three inequivalent $\mathbf{M}$ points rather than only one. In addition to the geometric ground-state energy signature, the momentum separation of its degenerate ground-state total momenta is exactly given by the momenta $\mathbf{M}_i$. In any case, a more reliable hallmark of the $\mathbf{M}$-WC is found in \cfig{fig:observables_nu_smaller}(b), where clear, distinctive Bragg peaks in the charge structure factor are present for all three order momenta $\mathbf{M}_i$. The finite-size extrapolation in \cfig{fig:observables_nu_smaller}(c) assures the prevalence of the $\mathbf{M}$-WC in the TDL. 

Considering the smaller filling of ${\nu=1/7}$, we focus on the geometric property of $C_6$ rotational symmetry. Figure \ref{fig:observables_nu_smaller}(d) highlights the lowered energy of larger clusters that are at least $C_6$ symmetric. On such lattices, the momentum-space spectrum shown in \capp{sec:derivation_Sq} in \cfig{fig:spectrum_ktot_nu_1_over_7}(a) or \cfig{fig:spectrum_ktot_nu_1_over_7}(b) displays a 14- or sevenfold ground-state degeneracy of orbitals separated by the six WC momenta of each $C_6$-WC class. The structure factor in \cfig{fig:observables_nu_smaller}(e) again exhibits the pronounced pattern of a $C_6$-WC, albeit the peak values of the two WC orientations on the cluster \nameref{cl:7.0.0.7} are slightly different in magnitude. This reflects the lack of a microscopic $C_2$ symmetry due to the hBN substrate, consistent with the minor energetic splitting of the ground-states depicted in the inset of \cfig{fig:spectrum_ktot_nu_1_over_7}(a). Finally, we average the order parameters at all $\mathbf{q}^\ast$ realizations to account for the splitting into two groups of peaks on the $D_6$ symmetric grid and perform a finite-size extrapolation. Although the small number of data points demands the final value of the regression to be taken with a grain of salt, the remnant normalized $C_6$-WC structure factor in the TDL is of the same order as for the $\mathbf{M}$-WC and the $\mathbf{K}$-CDW.

\subsection{Second hierarchy FCI at ${\nu=2/5}$ \label{sec:nu_2_over_5}}

\begin{figure*}[htp]
	\centering
	\includegraphics[width=0.9\textwidth]{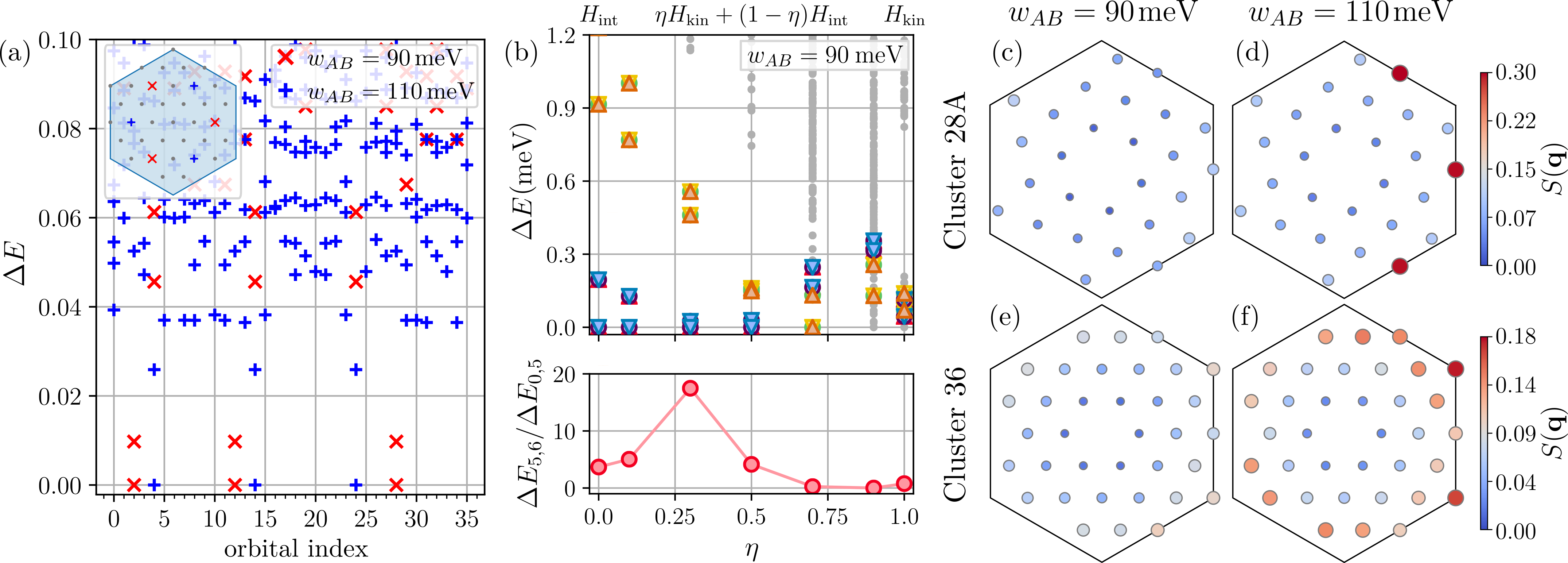}
	\caption{(a) Many-body spectrum at ${\nu=1/2}$ of the cluster \nameref{cl:6.0.0.6} with ground-state orbitals marked in the inset MBZ. The quasi-sixfold degenerate ground-state manifold of (a) at ${w_{AB}=\SI{90}{meV}}$ is stabilized by the single-particle dispersion in (b), as indicated by the increased gap to ground-state splitting ratio $\Delta E_{5,6}/\Delta E_{0,5}$. The ground-states at ${\eta=0}$ are marked in violet, red and blue while the next higher set of states is green, yellow and orange. (c--f) The measurement of $S(\mathbf{q})$ again signals an increased charge order tendency at ${w_{AB}=\SI{110}{meV}}$, although no universal order momentum could be identified and the peaks are less pronounced than for ${\nu \leq 1/3}$. \label{fig:observables_nu_1_over_2}}
\end{figure*}

With regard to valence band fillings above ${\nu=1/3}$, a potentially interesting filling fraction to study in more detail is ${\nu=2/5}$ as it is not only a candidate for the realization of a hierarchy FCI state \cite{Laeuchli2013,Liu2013} but was also found to exhibit insulating behavior in related TMD moir\'{e} heterostructures \cite{Xu2020}. A first look at the low-energy spectra, presented in \cfig{fig:observables_nu_2_over_5}(a,b), reveals manifest differences between the two considered hopping parameter values. While the five ground-states at ${w_{AB}=\SI{90}{meV}}$ agree with the ${\nu=2/5}$ FCI heuristics, the distribution of eigenvalues at ${w_{AB}=\SI{110}{meV}}$ is less obvious in its interpretation. The density correlation measurements of \cfig{fig:observables_nu_2_over_5}(c,d) suggest that the charge order tendency is once more increasingly pronounced at ${w_{AB}=\SI{110}{meV}}$ as opposed to ${w_{AB}=\SI{90}{meV}}$, although the sharpness of the peaks in $S(\mathbf{q})$ is significantly reduced compared to the results at ${\nu=1/3}$. Making use of the understanding acquired in \csec{sec:interplay}, we can further probe the nature of the ground-state via the introduction of the valence band dispersion. In accordance with preceding findings, we observe that the potential FCI ground-state manifold is stabilized by $H_{\text{kin}}$ via an increase of the excitation gap to ground-state splitting ratio until ${\eta \simeq 0.5}$, whereas the spectrum at ${w_{AB}=\SI{110}{meV}}$ collapses monotonically (not shown). We use the optimal convex combination for the FCI to perform the insertion of magnetic flux quanta in \cfig{fig:observables_nu_2_over_5}(e) and find that the ground-states exhibit the required spectral flow until ${\Phi_1/2 \pi=5}$. On the other hand, if we flip the single-particle dispersion and thus mimic the situation in the conduction band, at ${w_{AB}=\SI{110}{meV}}$ a series of 15 states separates from energetically higher states until ${\eta \simeq 0.7}$. This profit of Berry curvature avoidance is consistent with a tendency for charge order and what is more, the developing 15-fold degeneracy matches the expected degree for the charge pattern proposed in \cref{Xu2020} to explain the ${\nu=2/5}$ insulating state. Nevertheless, the precise real-space pattern could not be confirmed within the scope of this work, not least due to the lack of a numerically accessible larger symmetric cluster that supports the suggested pattern. Simulations on less symmetric discretizations with $N_k=30$, $35$, $40$ could not be found to clarify the situation at ${w_{AB}=\SI{110}{meV}}$, while they did affirm the prevalence of the FCI at ${w_{AB}=\SI{90}{meV}}$. We also analyzed the situation at ${\nu=1/5}$ towards the possibility of FQH-like order, however, despite some promising signatures in the location and degeneracy of ground-state orbitals, the evidence did not sustain across multiple cluster sizes.

\subsection{Numerical results for half filling\label{sec:nu_1_over_2}}

Finally, we present results at half filling  ${\nu=1/2}$. This is of particular interest in the FQH context since the investigation of a spin-polarized half filled Landau levels has produced a number of exciting theoretical proposals, such as the composite fermion Fermi sea~\cite{Halperin1993}, or the Moore-Read FQH state which hosts non-abelian Ising anyons~\cite{Moore1991}, or variants of charge ordered phases~\cite{Moessner1996}. Let us note, that here our numerical results turn out to be more ambiguous than the previously discussed fillings and the conclusive identification of the ground-state nature needs to be left to future work.

\begin{figure}[b]
	\centering
	\includegraphics[width=.9\linewidth]{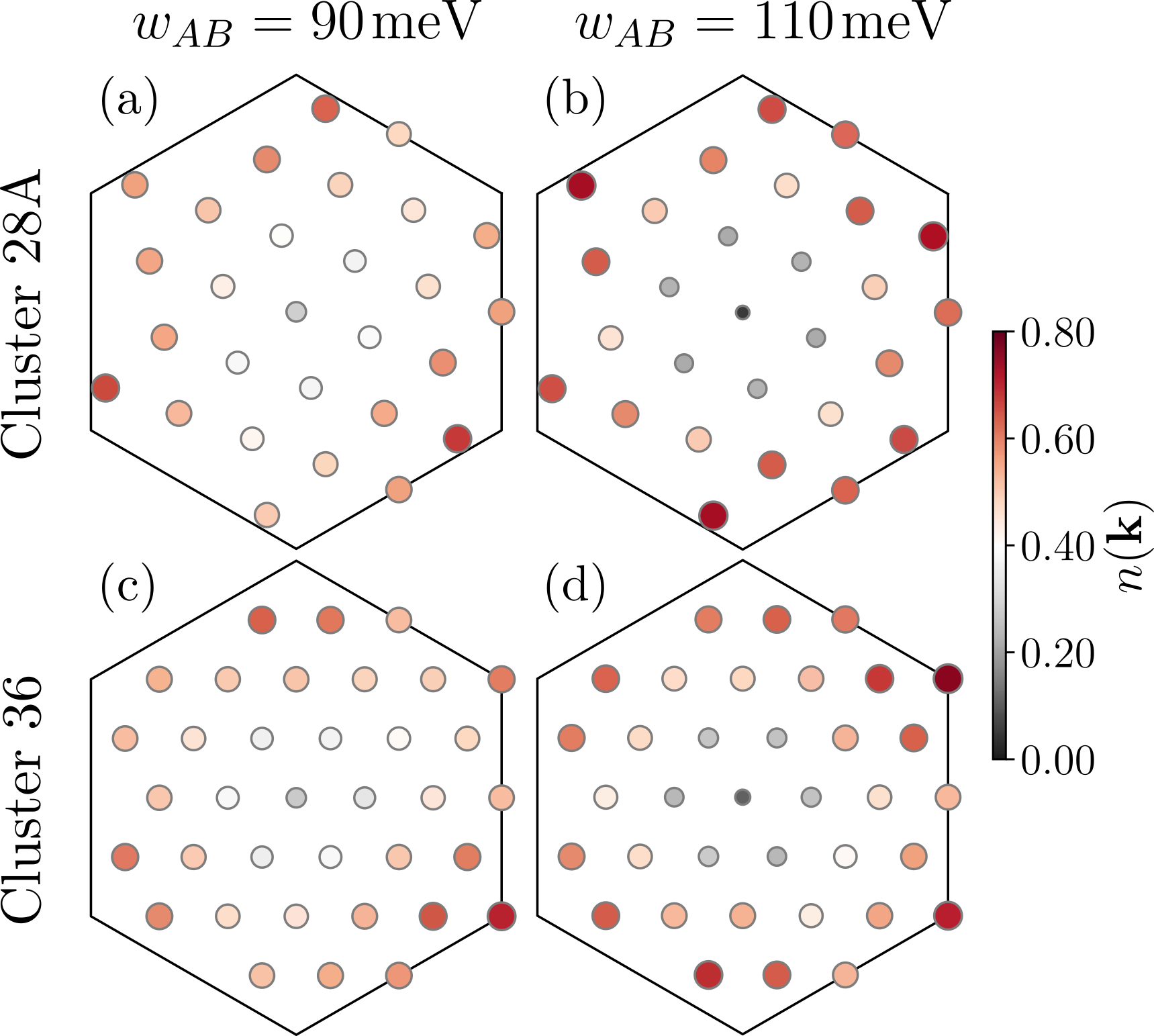}
	\caption{Ground-state orbital occupation of the clusters \nameref{cl:6.0.0.6} and \nameref{cl:2.4.6.-2} at ${\nu=1/2}$. Similar to \cfig{fig:occupation_nu_1_over_3}, ${w_{AB}=\SI{90}{meV}}$ leads to a more uniform occupation across the whole MBZ. \label{fig:nu_1_over_2_occupation}}
\end{figure}

\begin{figure}[b]
	\centering
	\includegraphics[width=.95\linewidth]{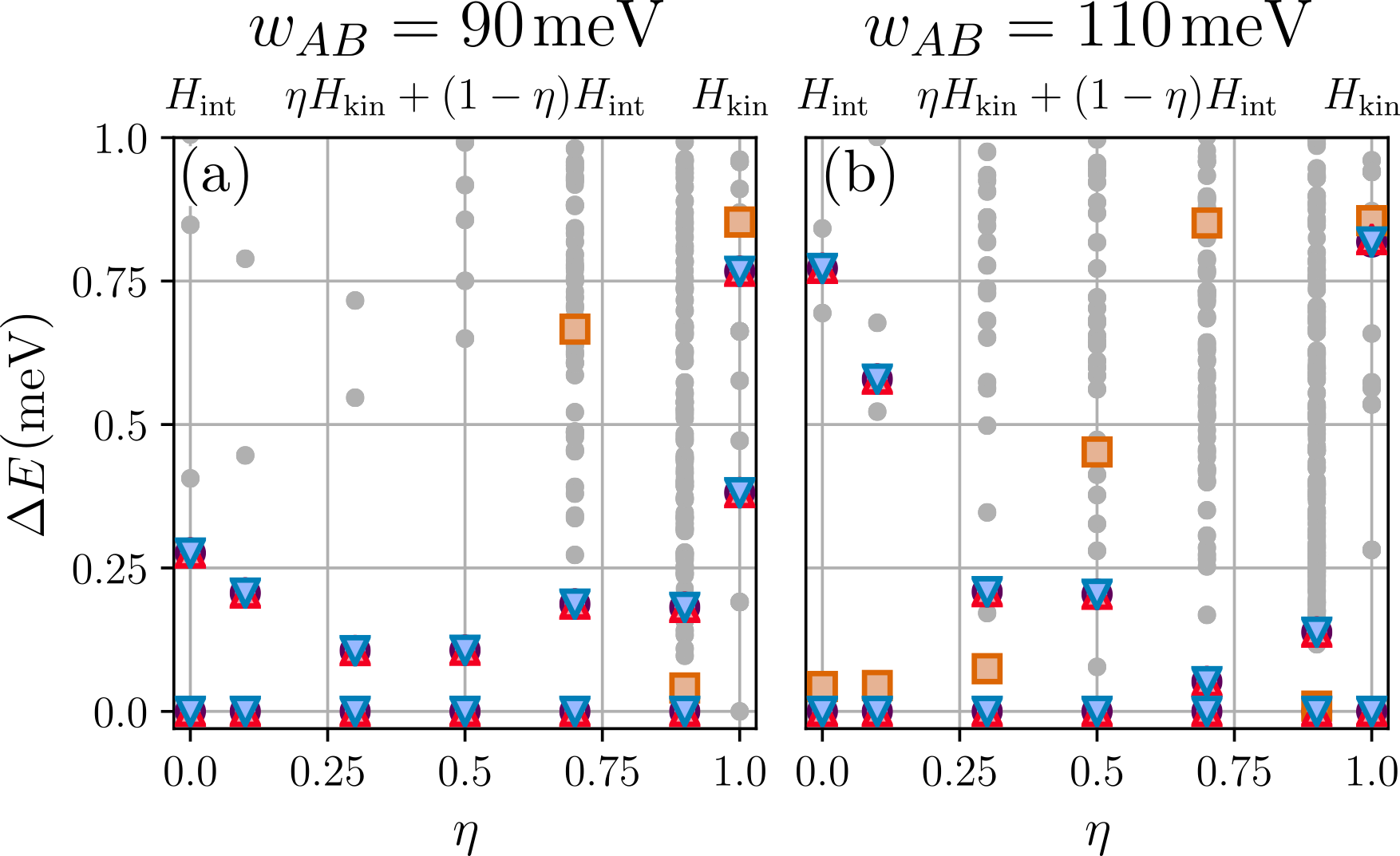}
	\caption{Introduction of a finite single-particle dispersion on the cluster \nameref{cl:2.4.6.-2} at ${\nu=1/2}$. The lowest two states at each of the $\mathbf{M}$ points are marked in violet, red and blue, while the energetically minimal one at $\mathbf{\Gamma}$ is orange.}
	\label{fig:convex_nu_1_over_2}
\end{figure}

The most intriguing signatures in our data are the imminent double-degeneracies of three momentum orbitals on clusters \nameref{cl:6.0.0.6} and \nameref{cl:2.4.6.-2} at ${w_{AB}=\SI{90}{meV}}$ in \cfig{fig:observables_nu_1_over_2}(a) as well as \cfig{fig:spectrum_ktot_nu_1_over_2} in \capp{sec:spectra}, which are reminiscent of the sixfold degenerate ${\nu=1/2}$ Pfaffian state \cite{Wen1993}. Similar to the FCI, such a FQH-like state intimately relies on the Chern character of the band in order to facilitate the formation of what is understood to be pairs of composite fermions \cite{Read2000}. We attempted an analysis of the Pfaffian orbital heuristics demanding two particles in four consecutive orbitals~\cite{Seidel2006, Bergholtz2006, Bernevig2008, Regnault2011, Wen2008a, Wen2008b}, but we obtained inconclusive results. While the ground-state COM orbitals on the cluster \nameref{cl:2.4.6.-2} at ${w_{AB}=\SI{90}{meV}}$ are consistent with the patterns "1010" and "0101" being realized in both momentum-loop directions, the same does not apply on \nameref{cl:6.0.0.6}. The observed cross-cluster variability of spectral features may be related to the differences in their topological extent, which was found to have a profound impact on the ground-state splitting of FCIs in \cref{Laeuchli2013} and might be the reason why certain orbital patterns are a priori suppressed. Also, the pure two-body nature of the interaction may be insufficient to stabilize a Pfaffian phase in this model.

Concerning the possibility a Fermi-liquid-like state driven by the IHD, an analysis of the generated Fermi surface yields a threefold degeneracy for both clusters \nameref{cl:2.4.6.-2} and \nameref{cl:6.0.0.6}, with COM orbitals at the $\mathbf{M}$ points, except for \nameref{cl:6.0.0.6} at ${w_{AB}=\SI{90}{meV}}$, where they are located slightly off the border of the MBZ. However, apart from the (partial) lack of agreement with the ground-state COM orbitals calculated by ED, the relatively miniscule energetic advantage of these configurations in the purely IHD-driven picture in conjunction with the absence of a clear Fermi surface in the orbital occupation $n(\mathbf{k})$ over $-E_h(\mathbf{k})$ at this filling, discussed in \cref{Abouelkomsan2020}, suggests more involved interaction effects beyond mere energetic preferences of the induced single-hole dispersion. The structure factor in \cfig{fig:observables_nu_1_over_2}(c--f) as well as the orbital occupations presented in \cfig{fig:nu_1_over_2_occupation} reveal comparable features to the ${\nu=1/3}$ case. The distribution of $n(\mathbf{k})$ is shifted towards the border of the MBZ, where the situation at $\SI{90}{meV}$ is once more smoother than at $\SI{110}{meV}$. Similarly, signatures in $S(\mathbf{q})$ signal an increased charge order tendency for $\SI{110}{meV}$ while such indications are suppressed at $\SI{90}{meV}$. Both of these observables were measured for the energetically lower and the higher lying state at the ground-state COM momenta. The results coincide qualitatively and quantitatively up to order $O(10^{-2})$. Although we find peaks in the structure factors at or in proximity to the $\mathbf{M}$ and $\mathbf{K}$ points for multiple clusters, the high degree of variability for this filling prohibits a stable finite-size extrapolation.
  
Similar to \csec{sec:nu_2_over_5}, we once more make use of the acquired understanding that a finite valence band dispersion supports effects from the inhomogeneous Berry curvature in order to investigate the nature of the sixfold degeneracy at ${w_{AB}=\SI{90}{meV}}$ and contrast it with the behavior at ${w_{AB}=\SI{110}{meV}}$. According to the top panel of \cfig{fig:observables_nu_1_over_2}(b), the effect of $H_{\text{kin}}$ at $\SI{90}{meV}$ is not as apparent as for ${\nu=1/3}$. Although the ground-state splitting (${\Delta E_{0,5}}$) decreases until ${\eta=0.3\mbox{--}0.5}$, the gap to the first excited state (${\Delta E_{5,6}}$) also decreases. By comparing the two quantities, the bottom panel of \cfig{fig:observables_nu_1_over_2}(b) demonstrates a substantial improvement of the excitation gap on the energy scale of the ground-state manifold. This observation is corroborated by \cfig{fig:convex_nu_1_over_2}, where the quality of the ground-state manifold is once more enhanced at ${w_{AB}=\SI{90}{meV}}$ up to ${\eta=0.5}$, while the order at ${w_{AB}=\SI{110}{meV}}$ is disfavored by the kinetic terms of the Hamiltonian. What is more, \cfig{fig:convex_nu_1_over_2}(b) at $\eta=0.7$ suggests the realization of a situation akin to ${w_{AB}=\SI{90}{meV}}$ for ${w_{AB}=\SI{110}{meV}}$, where the three $\mathbf{M}$ orbitals become almost doubly-degenerate.

The fact that a similar stabilization procedure to the ${1/3}$- and ${2/5}$-FCI applies for this configuration, hints at the quantum Hall-like nature of the phase at ${w_{AB}=\SI{90}{meV}}$. In addition, the appearance of such signatures at ${w_{AB}=\SI{110}{meV}}$ with an increased valence band dispersion is compatible with the pronounced peak of the Berry curvature at $\mathbf{\Gamma}$ for this hopping parameter. At the same time, the stabilization with $\eta$ provides further evidence against a Fermi-liquid driven by the IHD, since the valence $H_{\text{kin}}$ acts opposite to the preferences of $-E_h(\mathbf{k})$. 

At a filling fraction of $1/2$, another well known contender for the ground-state phase in a Landau level setting is the composite fermion liquid \cite{Halperin1993, Geraedts2016}. Since this is a metallic state, its Fermi surface may be responsible for the variable degeneracy of the ground-states on different clusters and, additionally, it may also profit from an increased importance of the Chern character of the band by altering $\eta$. Nevertheless, the impact of broken time-reversal and particle-hole symmetries in this model remain to be understood prior to a discussion on a more rigorous level. To sum up, although the designation of definitive ground-state orders for half filling would be too speculative based on the available data, our results contain crucial indications of the phases' nature.

\section{Drafting of a tentative phase diagram \label{sec:composition}}

\begin{figure}[htp]
	\includegraphics[width=0.9\linewidth]{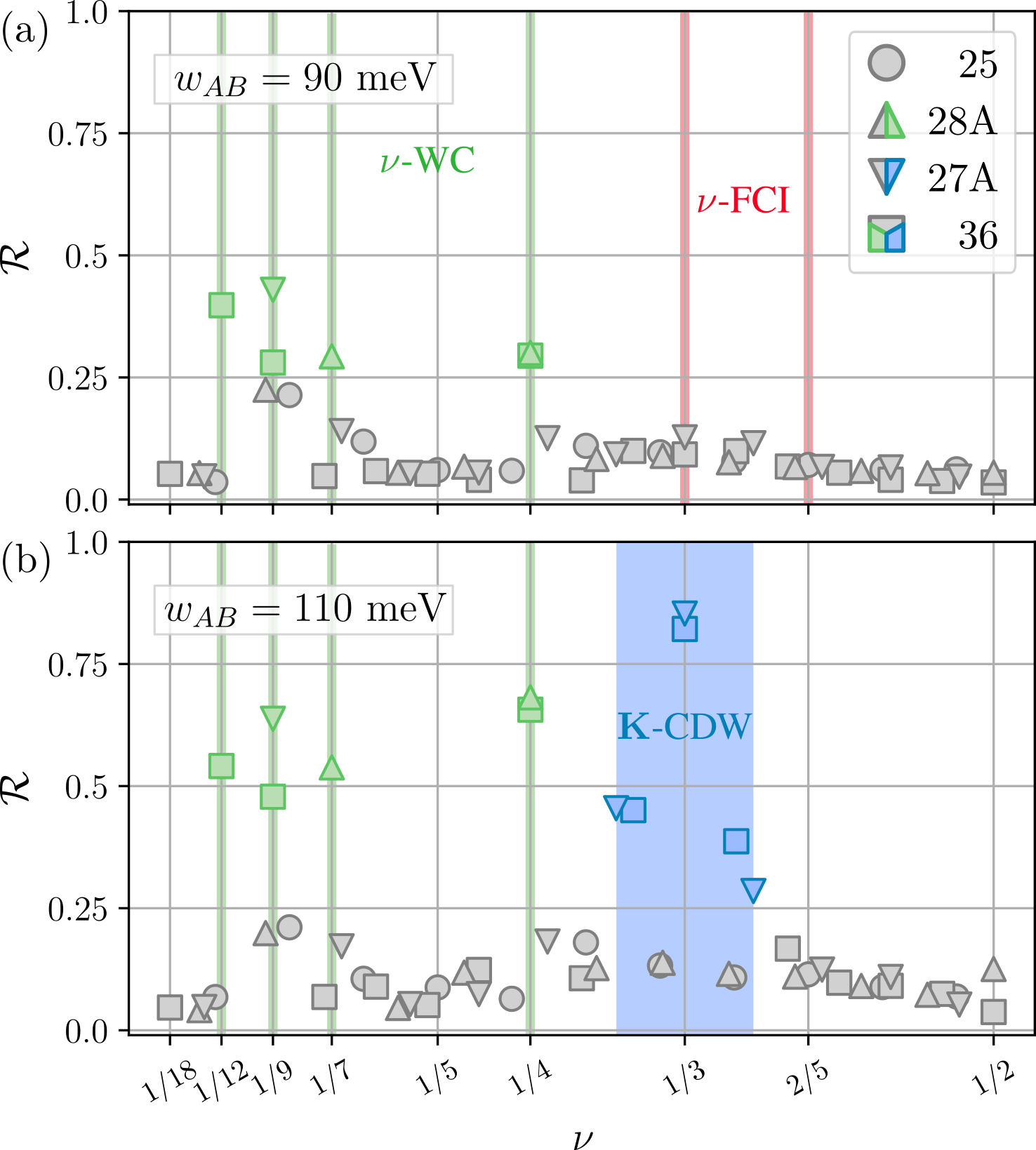}
	\caption{Structure factor sharpness $\mathcal{R}$ and identified regions of correlated phases over the scanned filling range $\nu$ for both interlayer hopping amplitudes. The appearance of significant signals only at the commensurate fillings ${\nu=1/12}$, $1/9$, $1/7$, $1/4$ suggests WC-type order (green) for both $w_{AB}$, whereas the extended region near ${\nu=1/3}$ supports the formation of a more robust $\mathbf{K}$-CDW (blue) at ${w_{AB}=\SI{110}{meV}}$. Evidence for the FCI (red) at ${w_{AB}=\SI{90}{meV}}$ was found for ${\nu=1/3}$ as well as ${\nu=2/5}$. Data points below $\mathcal{R}=0.25$ are marked in grey. Different symbols represent data from specific clusters. More details on the used clusters is found in \ctab{tab:clusters}. \label{fig:R_composed}}
\end{figure}

The abundance of data presented throughout \csec{sec:nu_canonical} calls for a more condensed graphical representation of the conclusive findings. Furthermore, the robustness of the different charge order patterns against a density deviation from their nominal filling has not been explored yet. In order to address both of these issues, we plot the structure factor ratio ${\mathcal{R} = S(\mathbf{q}^{\ast})/[\bar{S}(\mathbf{q}^{\ast} + \delta \mathbf{q})N_k]}$ normalized to the system size for multiple clusters at fillings ranging ${2\leq N_e \leq N_k/2}$ and overlay it with the unambiguously identified correlated phases in \cfig{fig:R_composed}. Here ${\bar{S}(\mathbf{q}^{\ast} + \delta \mathbf{q})}$ denotes the average contribution of momenta closest to $\mathbf{q}^{\ast}$, which are not related to $\mathbf{q}^{\ast}$ by a $C_6$ symmetry operation. In order to counteract band-projection artifacts and for added robustness in degenerate situations, we average all $C_6$ related contributions prior to the computation of $\mathcal{R}$. As $\mathcal{R}$ measures the sharpness of the peak in the structure factor, it is related to the correlation length of the charge density in real-space. Large $\mathcal{R}$ suggest pronounced long-range order, whereas small values of $\mathcal{R}$ argue against the presence of a distinct real-space order pattern. 
We choose to restrict to the pure interaction case of ${\eta=0}$ in order to avoid any bias on the charge order signatures stemming from the effects discussed in \csec{sec:interplay} and to keep the results applicable to the conduction band. Comparing \cfig{fig:R_composed}(a) with \cfig{fig:R_composed}(b), we can immediately tell that the two parameter values appear to result in similar physics at low fillings ${\nu\lesssim 1/4}$, while they differ substantially for a larger number of electrons per orbital ${\nu\gtrsim 1/3}$. The series of WC-like charge order continues to even smaller fillings of ${\nu=1/9}$ and ${\nu=1/12}$, with appropriate spectral features but also substantial peaks in $\mathcal{R}$. Similar to the ones at ${\nu=1/4}$, $1/7$, the abrupt reduction of their respective charge density correlation signature indicates that these charge orders manifest only at their corresponding commensurate filling fraction - highlighting their crystalline character. On the other hand, the $\mathbf{K}$-CDW near ${\nu\simeq1/3}$ at ${w_{AB}=\SI{110}{meV}}$ appears to be robust against the introduction or removal of a few additional electrons, making it the preferred order tendency across a whole range of fillings, featuring \emph{true} CDW character. Despite the composition of data from multiple clusters with different prime factorizations, the lack of pronounced charge order peaks slightly off the commensurate fillings ${\nu=1/4}$, $1/7$, $1/9$, $1/12$ may also be rooted in the relatively coarse resolution of \cfig{fig:R_composed} or the chosen metric $\mathcal{R}$ and a more CDW-like character may emerge in larger clusters. In addition to the symmetry breaking WC phases at low electron densities, the ${w_{AB}=\SI{90}{meV}}$ system also features topological FCI states at fillings ${\nu=1/3}$ as well as ${\nu=2/5}$. The absence of such states at ${\nu\lesssim 1/4}$, e.g. at ${\nu=1/7}$, $1/9$, affirms the intuition gathered throughout \csec{sec:interplay}, where the Coulomb interaction structure is found to generally prefer the arrangement of electrons at the border of the MBZ. At small fillings the electrons hence almost completely avoid the Berry curvature, rendering the situation qualitatively identical to ${w_{AB}=\SI{110}{meV}}$. Concerning the (partially) inconclusive filling fractions of ${\nu=2/5}$ and ${\nu=1/2}$, \cfig{fig:R_composed} attenuates the role of charge order in comparison to other, more pronounced situations.

A comparison of \cfig{fig:R_composed}(b) with very recent experimental results for the TMD based moir\'{e} system in \cref{Xu2020} suggests remarkable similarities with the TBLG/hBN structure discussed here. At ${w_{AB}=\SI{110}{meV}}$, where the Chern character of the TBLG/hBN flat band is found to be subordinate, the experimental findings and proposed real-space order patterns at ${\nu=1/7}$, $1/4$ and ${\nu=1/3}$ coincide with our theoretical predictions.

\section{Conclusion}

We performed an extensive exact diagonalization study of the single-band-projected TBLG/hBN many-body model at fractional fillings in the momentum-space basis. For a band filling of ${\nu=1/3}$, we showed that the screened Coulomb interaction between electrons enables the formation of both a topological FCI but also a geometry sensitive CDW state. For ${w_{AB}=\SI{90}{meV}}$ and upon neglecting the single-particle dispersion, we agree with \cref{Abouelkomsan2020} and \cref{Repellin2020} on the FCI nature of the ground-state in the valence and conduction bands. However, as the interlayer hopping amplitude is increased to $\SI{110}{meV}$, we obtained solid evidence for a CDW with Dirac point order momentum that spontaneously breaks moir\'{e} translational symmetry and triples the unit cell. Signatures in the spectra and the structure factor point to the competition of these correlated insulating phases at ${w_{AB}=\SI{90}{meV}}$, while ${w_{AB}=\SI{110}{meV}}$ clearly favors the $\mathbf{K}$-CDW, even for a fraction of the original screening length. This competition is further highlighted upon including the realistic kinetic energy contribution. While the opposing energetic preferences of the single-particle and the interaction induced hole dispersion in the valence band at ${w_{AB}=\SI{90}{meV}}$ lead to an FCI sweet spot where the electron density is smoothed across the MBZ, the flipped dispersion of the conduction band instead reinforces the tendency to occupy orbitals at the boundary and thus suppresses the FCI state in favor of the CDW. At ${w_{AB}=\SI{110}{meV}}$, the kinetic energy gradually penalizes the CDW state energetically until $H_{\text{kin}}$ becomes the dominant energy scale for the ground-state. The behavior at the complementary ${\nu=2/3}$ filling can be well explained by the situation at filling $1/3$ with twice the amount of electrons to accommodate in the MBZ. Further investigations of possible charge order at the next smaller commensurate fillings ${\nu=1/4}$, $1/7$, corresponding to a four- or sevenfold extension of the unit cell, lead to the conclusion that such a symmetry breaking correlated insulator may quite generically form in this model. Apart from the evidence for the formation of WCs, we corroborate the analogy to Landau levels beyond the ${\nu=1/3}$ state by demonstrating convincing signatures of a ${\nu=2/5}$-FCI at ${w_{AB}=\SI{90}{meV}}$. The situation at half filling turned out to be much more involved and could not be resolved unambiguously on the available cluster sizes. Nevertheless, we found qualitative similarities in the observables compared to other filling fractions, which together might contribute to a more comprehensive understanding in the future. The wealth of conclusive results is finally condensed and put into perspective in a tentative phase diagram for the filling dependence of order tendencies in TBLG/hBN, which, among other things reveals the $\mathbf{K}$-CDW character near ${\nu=1/3}$, while charge order throughout the commensurate density series ${\nu=1/4}$, $1/7$, $1/9$, $1/12$ is of WC-type, i.e.~locked to the lattice at the corresponding commensurate densities.

We furthermore developed intuition on what microscopic mechanism drives the (de-) stabilization of the two phases: The interplay of the induced hole dispersion and kinetic energy, which essentially determine the electron density distribution, with the effective magnetic field due to the Berry curvature appears to be the fundamental reason the system favors one correlated phase over the other for very similar band parameters.

Our results thus promote the translational symmetry breaking charge-density-wave to a probable order tendency for the real moir\'{e} system. 
Our findings highlight the system's sensitivity to microscopic model parameters even in the idealized situation of our treatment. This is in accordance with the issue of strong sample-to-sample dependence in experiments, where twist angle homogeneity, strain or pressure can directly affect the degree of interlayer orbital overlap. The recent evidence for $\mathbf{K}$- and (stripe) $\mathbf{M}$-CDWs in \cref{Xie2020a} for unaligned TBLG, at an electron filling roughly corresponding to ${\nu=1/4}$ in our flavor-polarized model, affirms the relevance of our results that charge order represent a general order tendency across multiple filling fractions to the physics of pure TBLG. The implications of our work are further extended by the agreement with recent experimental findings for a TMD based heterostructure in \cref{Xu2020}, suggesting a remarkable resemblance of these moir\'{e} systems for certain parameter regions.

\section*{Acknowledgments}

We thank M.~S.~Scheurer for valuable discussions and comments. Moreover, we are grateful to  A.~Wietek and M.~Schuler for providing the QuantiPy package, which simplified the incorporation of diverse simulation clusters. We acknowledge support by the Austrian Science Fund FWF within the DK-ALM (W1259-N27). The computational results presented have been achieved in part using the Vienna Scientific Cluster (VSC).

\FloatBarrier
\begin{appendix}

\section{Additional many-body spectra \label{sec:spectra}}

\begin{figure}[htp]
	\includegraphics[width=\linewidth]{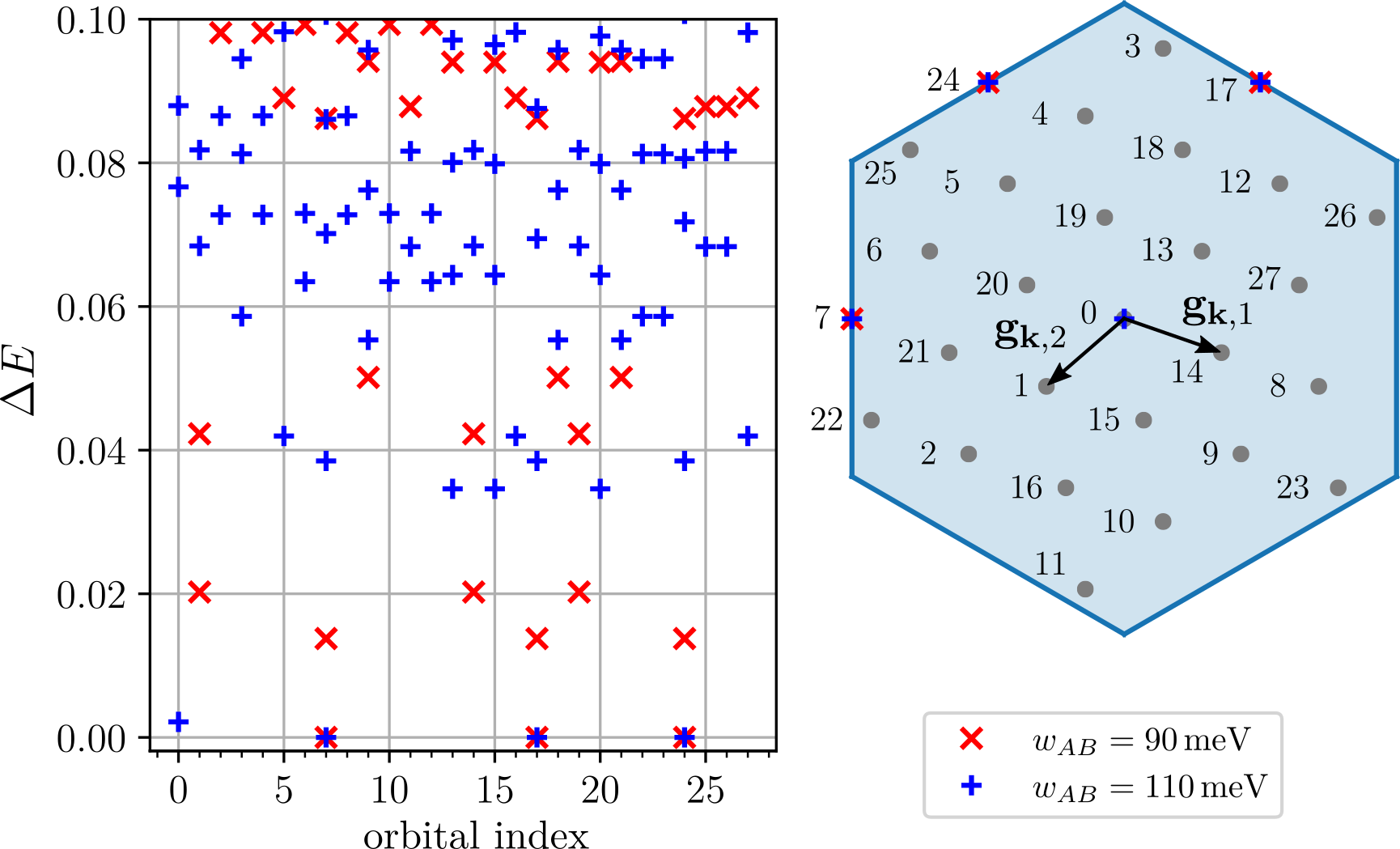}
	\caption{Spectrum and location of ground-state orbitals for the cluster \nameref{cl:2.4.6.-2} at ${\nu=1/2}$ filling. \label{fig:spectrum_ktot_nu_1_over_2}}
\end{figure}

\begin{figure}[htp]
	\includegraphics[width=\linewidth]{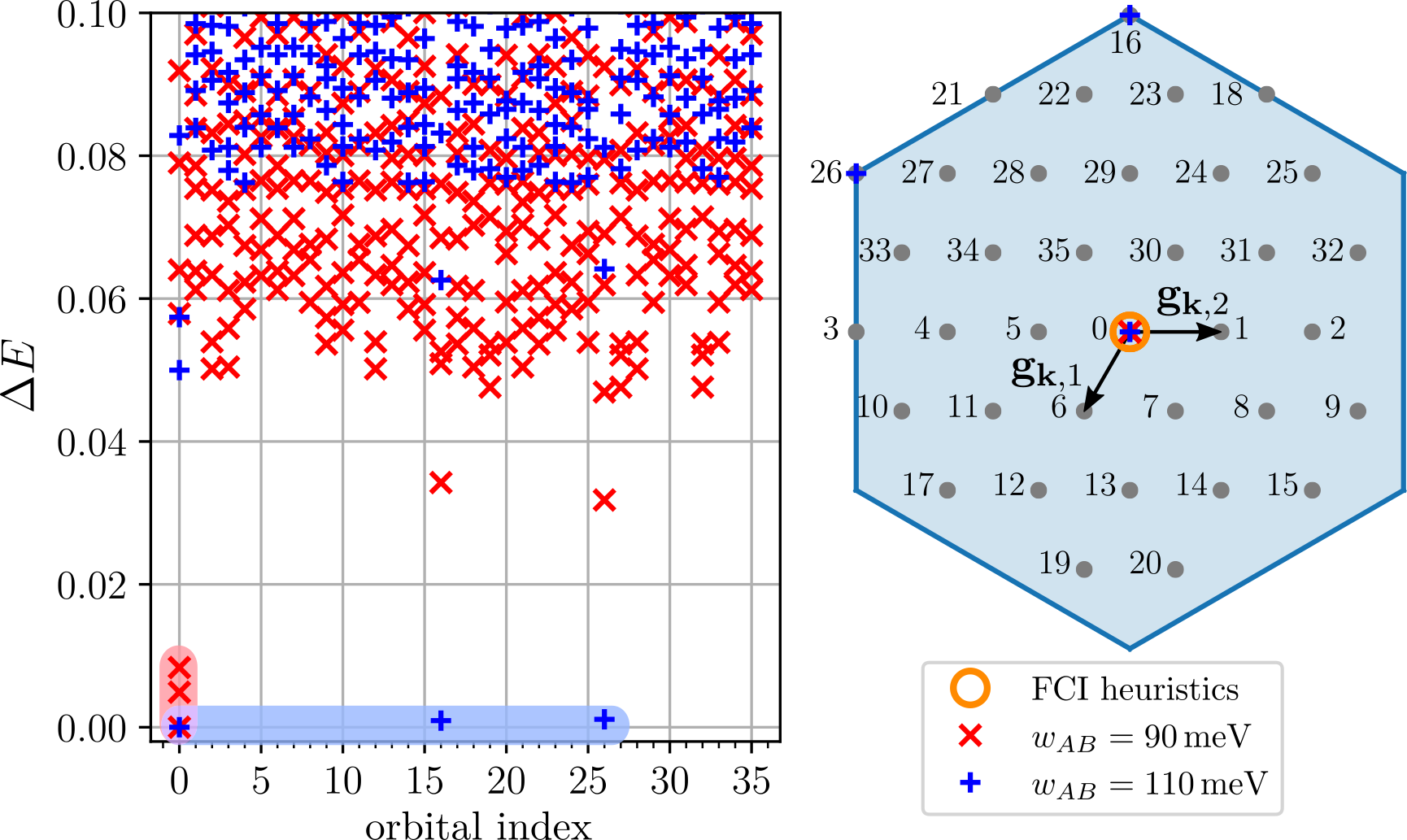}
	\caption{Spectrum and location of ground-state orbitals for the cluster \nameref{cl:6.0.0.6} at ${\nu=1/3}$ filling with $\lambda=L^{\text{M}}/6$. \label{fig:spectrum_ktot_nu_1_over_3_LM6th}}
\end{figure}

\begin{figure}[htp]
	\centering
	\includegraphics[width=.95\linewidth]{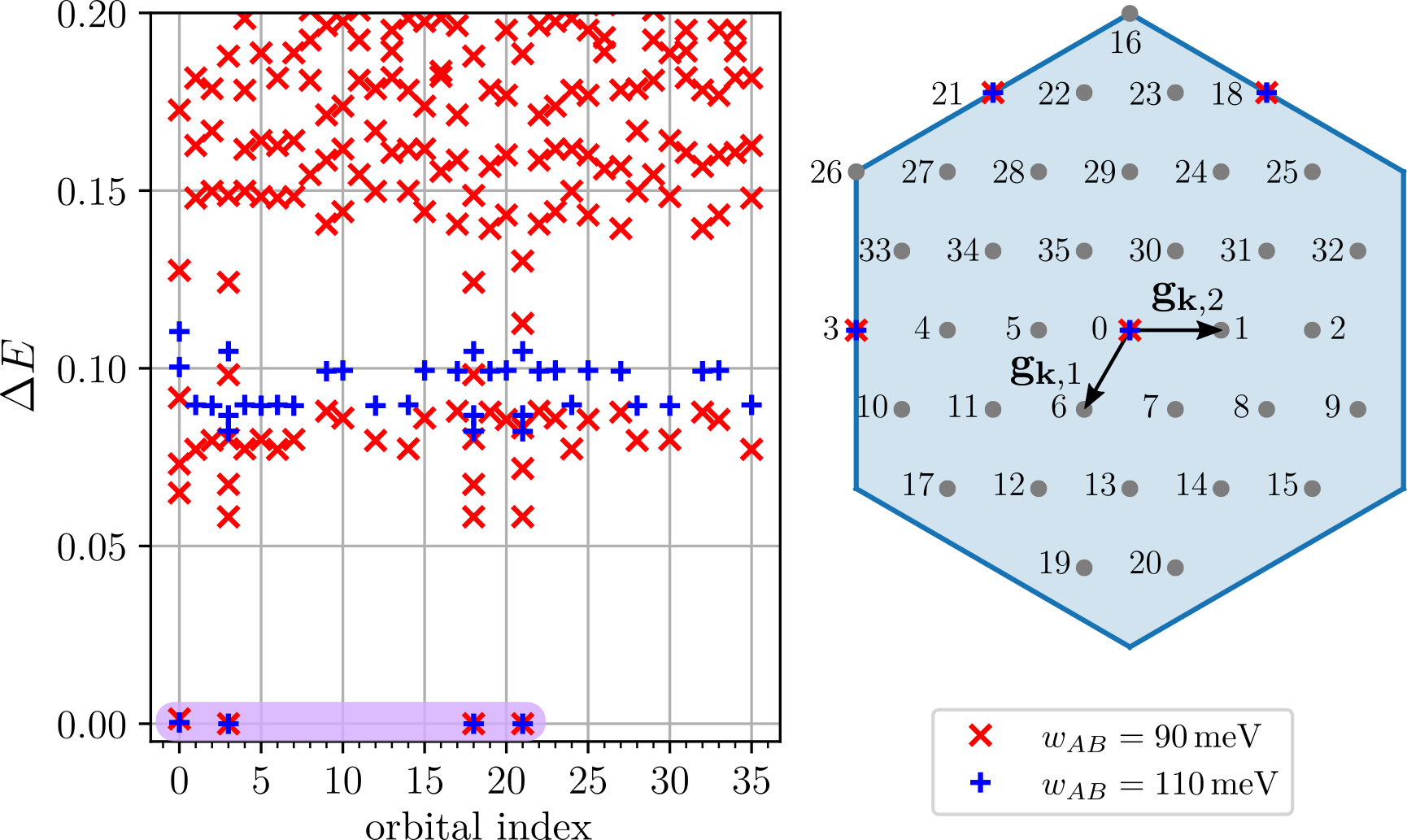}
	\caption{Spectrum and location of ground-state orbitals for the cluster \nameref{cl:6.0.0.6} at ${\nu=1/4}$ filling.}
	\label{fig:spectrum_ktot_nu_1_over_4}
\end{figure}

\begin{figure}[htp]
	\begin{minipage}[b]{.95\linewidth}
	\centering
		\begin{minipage}[b]{\linewidth}
			\includegraphics[width=\linewidth]{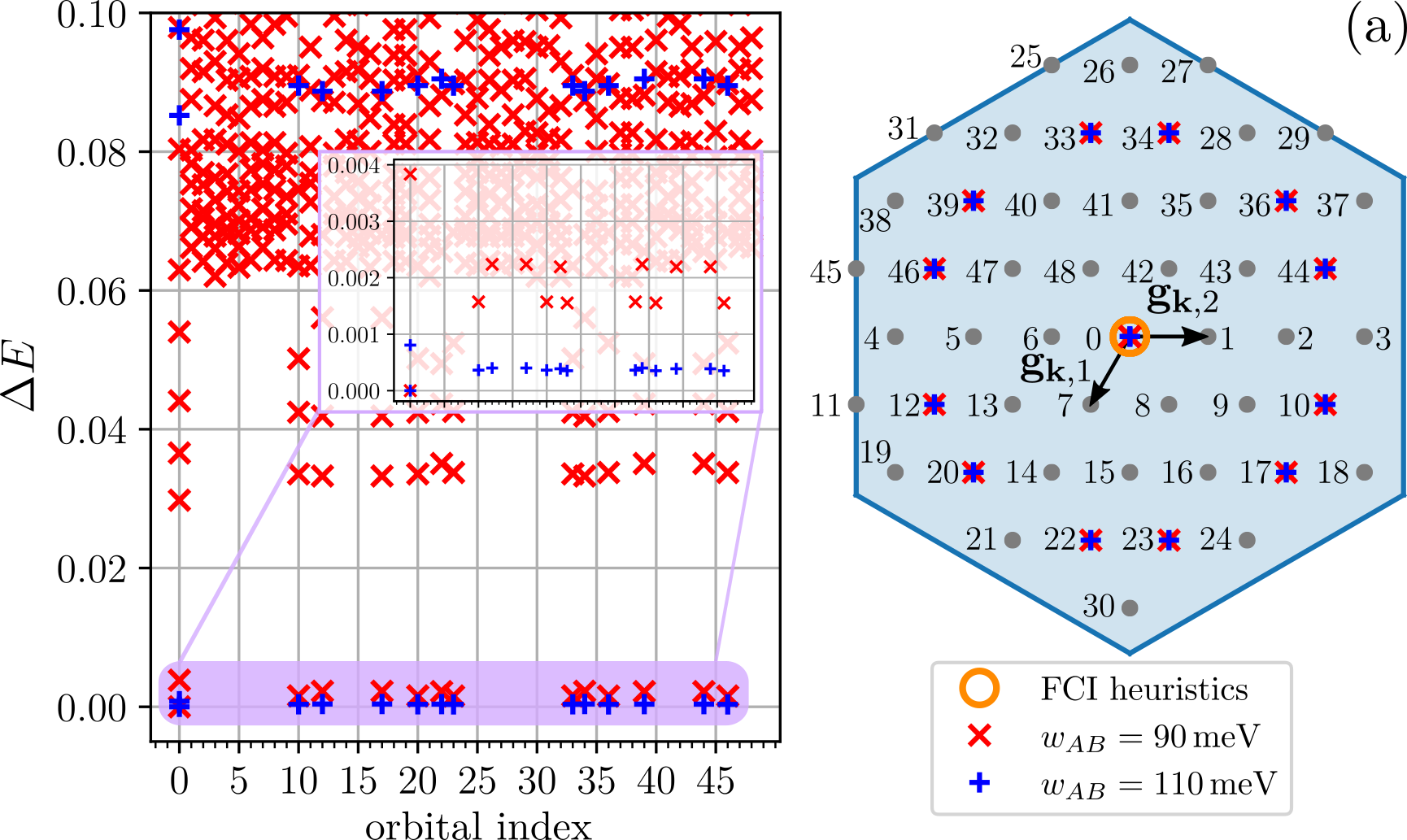}
		\end{minipage}q
		\begin{minipage}[b]{\linewidth}
			\includegraphics[width=\linewidth]{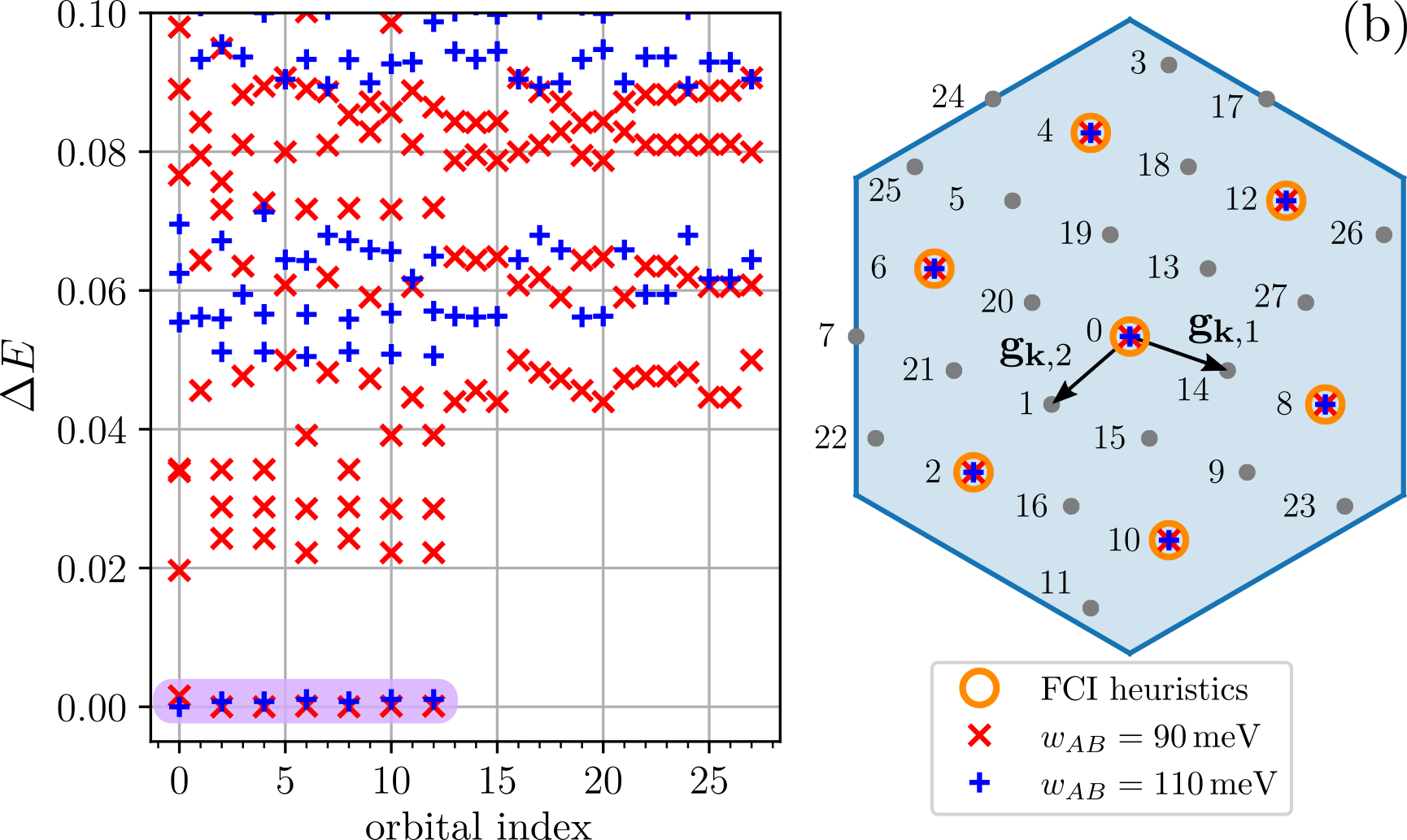}
		\end{minipage}
	\end{minipage}
	\caption{Spectrum and location of ground-state orbitals for the cluster (a) \nameref{cl:7.0.0.7} and (b) \nameref{cl:2.4.6.-2} at ${\nu=1/7}$ filling. The inset in (a) highlights the substrate induced splitting of the ground-state manifold into two $C_6$ related sets of orbitals.	\label{fig:spectrum_ktot_nu_1_over_7}}
\end{figure}

This section provides an overview of exemplary many-body spectra encountered in the ED study but not included in the main text. The identified ground-state manifolds are shaded in the color of the respective symbols. The displayed results include further quasi double-degeneracies at half filling in \cfig{fig:spectrum_ktot_nu_1_over_2} on the cluster \nameref{cl:2.4.6.-2} as well as the evidence for stable $\mathbf{K}$-CDW order at ${\nu=1/3}$ with the shorter screening length of ${\lambda=L^\text{M}/6}$ in \cfig{fig:spectrum_ktot_nu_1_over_3_LM6th}. The degeneracy and orbital separation in \cfig{fig:spectrum_ktot_nu_1_over_4} clearly indicate an $\mathbf{M}$-WC for both considered values of $w_{AB}$. Figure~\ref{fig:spectrum_ktot_nu_1_over_7}(a) highlights the possibility for two classes of a $C_6$-WC on clusters with $D_6$ symmetry like \nameref{cl:7.0.0.7}, leading to an approximate 14-fold ground-state degeneracy with a minor energetic splitting due to the substrate induced breaking of $C_2$. On the other hand, $C_6$ symmetric clusters similar to \nameref{cl:2.4.6.-2} in \cfig{fig:spectrum_ktot_nu_1_over_7}(b) can realize only a single variant of the translation symmetry breaking WC. The momenta of a given cluster are addressed by integers $k_1$, $k_2$, such that ${\mathbf{k} = k_1 \mathbf{g}_{\mathbf{k},1} + k_2 \mathbf{g}_{\mathbf{k},2}}$. The number of steps along $\mathbf{g}_{\mathbf{k},2}$ until the origin is reencountered is denoted by $N_2$ and is related to the topological length of \cref{Laeuchli2013}. For the displayed clusters \nameref{cl:5.0.0.5}, \nameref{cl:6.0.0.6}, \nameref{cl:7.0.0.7}, \nameref{cl:2.4.6.-2} we obtain ${N_2 = 5}$, $6$, $7$, $14$.

\section{Derivation of the band-projected structure factor \label{sec:derivation_Sq}}

The structure factor is generally defined as the Fourier transform of the static density-density correlation function 
\begin{equation}
	\begin{aligned}
		\chi_0(\mathbf{r}_i, \mathbf{r}_j) &= \left\langle \rho(\mathbf{r}_i) \rho(\mathbf{r}_j) \right\rangle \\
		&= \frac{1}{N_k} \sum_{\tilde{\mathbf{q}}} e^{i \tilde{\mathbf{q}} (\mathbf{r}_j - \mathbf{r}_i)} S(\tilde{\mathbf{q}})\,,
	\end{aligned}
\end{equation}
with ${S(\tilde{\mathbf{q}})=\frac{1}{N_k} \langle \rho(\tilde{\mathbf{q}}) \rho(-\tilde{\mathbf{q}}) \rangle}$ for general fermionic momentum-space density operators ${\rho(\tilde{\mathbf{q}}) = \sum_{\tilde{\mathbf{k}}} f^\dagger_{\tilde{\mathbf{k}}} f^{\pd}_{\tilde{\mathbf{k}} + \tilde{\mathbf{q}}}}$. In our notation, the momenta $\tilde{\mathbf{k}}$ and  $\tilde{\mathbf{q}}$ are located inside the Brillouin zone of ordinary graphene and thus have to be folded back onto ${\mathbf{k},\mathbf{q} \in \text{MBZ}}$ via ${\tilde{\mathbf{k}} = \mathbf{k} + \mathbf{G}}$ and ${\tilde{\mathbf{q}} = \mathbf{q} + \mathbf{G}}$. Since we are interested in the dominant correlations on the moir\'{e} scale, we restrict to the measurement of $S(\mathbf{q})$. This means we consider only momentum transfers $\mathbf{q}$ in the original MBZ and effectively drop the sum over $\mathbf{G}$, which would otherwise be introduced by the transformation to the band basis [see \ceqn{eq:matrix_elements}]. In the continuum model, the graphene second quantized operators are indexed by valley $\tau$, sublattice $X$ and momentum $\tilde{\mathbf{k}}$, which in the moir\'{e} band basis transforms to $\tau$, band $n$ and $\mathbf{k}$ as
\begin{align}
	\label{eq:band_transform}
	f^{\pd}_{\tau, X, \mathbf{k} + \mathbf{G}} = \sum_{n} u_{\tau,n;\mathbf{G},X}(\mathbf{k}) c^{\pd}_{\tau,n,\mathbf{k}}\,.
\end{align}
The eigenvector-components $u_{\tau,n;\mathbf{G},X}(\mathbf{k})$ are obtained from solving the single-particle eigenproblem and introduce additional form factors into the expression for the structure factor. Here it should be noted that the truncation to $\mathbf{G}=\mathbf{0}$ may give rise to slight quantitative discrepancies in the measurement, mostly for large $\mathbf{q}$ at the border of the MBZ, depending on what contributions at the boundary of the MBZ are taken into account. Nevertheless, these are only minor effects and what is more, the inclusion of $\mathbf{G} \neq \mathbf{0}$ contributions was found to reproduce the qualitative aspects of the results. As we consider only spinless fermions of a single band and valley and thus neglect band indices in \ceqn{eq:Sq_trafo} from line 2 onward, the transformation reads
\begin{equation}
	\label{eq:Sq_trafo}
	\begin{aligned}
		S(\mathbf{q})&=\frac{1}{N_k} \left\langle \sum_{X_1,\tilde{\mathbf{k}}_1} f^\dagger_{X_1,\tilde{\mathbf{k}}_1} f^{\pd}_{X_1,\tilde{\mathbf{k}}_1 + \mathbf{q}} \sum_{X_2,\tilde{\mathbf{k}}_2} f^\dagger_{X_2,\tilde{\mathbf{k}}_2} f^{\pd}_{X_2,\tilde{\mathbf{k}}_2 - \mathbf{q}} \right\rangle\\
		&= \frac{1}{N_k} \sum_{\mathbf{k}_1, \mathbf{k}_2} \Lambda_{\mathbf{k}_1}^{\mathbf{q}} \Lambda_{\mathbf{k}_2}^{-\mathbf{q}} \left\langle c_{\mathbf{k}_1}^{\dagger} c_{\mathbf{k}_1 + \mathbf{q}}^{\pd} c_{\mathbf{k}_2}^{\dagger} c_{\mathbf{k}_2 - \mathbf{q}}^{\pd} \right\rangle \\
		&= \frac{1}{N_k} 
		\begin{alignedat}[t]{1}
			\sum_{\mathbf{k}_1, \mathbf{k}_2}  \Lambda_{\mathbf{k}_1}^{\mathbf{q}} \Lambda_{\mathbf{k}_2}^{-\mathbf{q}} &\left[ \delta_{\mathbf{k}_1 + \mathbf{q}, \mathbf{k}_2} \left\langle c_{\mathbf{k}_1}^{\dagger} c_{\mathbf{k}_2 - \mathbf{q}}^{\pd} \right\rangle  - \right. \\
			&\left. \left\langle c_{\mathbf{k}_1}^{\dagger} c_{\mathbf{k}_2}^{\dagger} c_{\mathbf{k}_1 + \mathbf{q}}^{\pd} c_{\mathbf{k}_2 - \mathbf{q}}^{\pd} \right\rangle \right]
		\end{alignedat}\\
		&= \frac{1}{N_k} \begin{alignedat}[t]{1} 
			& \left[ \sum_{\mathbf{k}} \left|\Lambda_{\mathbf{k}}^{\mathbf{q}}\right|^2 n(\mathbf{k})  + \right. \\
			& \left. \sum_{\mathbf{k}_1, \mathbf{k}_2} \Lambda_{\mathbf{k}_1}^{\mathbf{q}} \Lambda_{\mathbf{k}_2}^{-\mathbf{q}} \left\langle c_{\mathbf{k}_1}^{\dagger} c_{\mathbf{k}_2}^{\dagger} c_{\mathbf{k}_2 - \mathbf{q}}^{\pd} c_{\mathbf{k}_1 + \mathbf{q}}^{\pd} \right\rangle \right]\,,
		\end{alignedat}
	\end{aligned}
\end{equation}
with $\Lambda_{\mathbf{k}}^{\mathbf{q}}$ again denoting the form factors introduced in \ceqn{eq:matrix_elements}.

\section{Used cluster geometries \label{sec:clusters}}

Table~\ref{tab:clusters} gives an overview of all the clusters used for performing ED. Each one has a distinct ID, which it is referred to by in the main text. The geometric properties of aspect ratio, number of realizations of high symmetry momenta and the point group are the basis for choosing a viable cluster in the first place but also guide the interpretation of numerical results. The torus spans the real-space simulation cell like ${\mathbf{T}_1 = a \mathbf{L}^\text{M}_1 + b \mathbf{L}^\text{M}_2}$ and ${\mathbf{T}_2 = c \mathbf{L}^\text{M}_1 + d \mathbf{L}^\text{M}_2}$, where $\mathbf{L}^\text{M}_i$ are the moir\'{e} lattice vectors. The momentum-space discretization $\mathbf{g}_{\mathbf{k},i}$ may then be derived as usual by finding the respective reciprocal vectors.

\begin{table}[t]
	\centering
	\caption{Overview of the cluster geometries used in this work. Depending on their realizations of high symmetry momenta $\mathbf{K}_{\pm}$, $\mathbf{M}$ and point symmetry group they may support different types of charge order.}
	\label{tab:clusters}
	\resizebox{\linewidth}{!}{
	\begin{tabular}{L{1cm}C{1cm}C{2cm}C{1cm}C{1cm}C{1cm}R{1cm}}
		\hline
		\hline
		\multirow{2}{*}{ID}\vspace{-.5em} & \multirow{2}{*}{$N_k$} & \multirow{2}{*}{\shortstack{Torus \\ $[[a,b],[c,d]]$}} & \multirow{2}{*}{\shortstack{Aspect \\ ratio}} &\multicolumn{2}{c}{Number of } & \multirow{2}{*}{\shortstack{Point\\group}}\\
		& & & & $\mathbf{K}_{\pm}$ & $\mathbf{M}$ & \\
		
		\hline
		\labelText{12A}{cl:2.2.2.-4} & 12 &  $ [[2,2],[2,-4]]$ & 1.00 &  1 &  3& $D_6$\\
		\labelText{12B}{cl:3.0.0.4} &12 &  $[[3,0],[0,4]]$ & 1.33 &  0 &  1& $C_2$\\
		\labelText{15A}{cl:1.3.4.-3} &15 &  $[[1,3],[4,-3]]$ & 1.00 &  0 &  0& $D_2$\\
		\labelText{15B}{cl:3.0.0.5} &15 &  $[[3,0],[0,5]]$ & 1.67 &  0 &  0& $C_2$\\
		\labelText{16}{cl:4.0.0.4} &16 &  $[[4,0],[0,4]]$ & 1.00 &  0 &  3 & $D_6$\\
		\labelText{18}{cl:3.0.0.6} &18 &  $[[3,0],[0,6] ]$ & 2.00 &  1 &  1& $D_2$\\
		\labelText{20A}{cl:2.2.4.-6} &20 &  $[[2,2],[4,-6] ]$ & 1.53 &  0 &  3& $D_2$\\
		\labelText{20B}{cl:2.-4.3.4} &20 &  $[[2,-4],[3,4] ]$ & 1.76 &  0 &  1& $D_2$\\
		\labelText{21A}{cl:1.4.5.-1} &21 &  $[[1,4],[5,-1] ]$ & 1.00 &  1 &  0& $C_6$\\
		\labelText{21B}{cl:3.0.0.7} &21 &  $[[3,0],[0,7] ]$ & 2.33 &  0 &  0& $C_2$\\
		\labelText{24A}{cl:1.4.5.-4} &24 &  $[[1,4],[5,-4] ]$ & 1.00 &  1 &  1 & $D_2$\\
		\labelText{24B}{cl:2.2.6.-6} &24 &  $[[2,2],[6,-6] ]$ & 1.73 &  1 &  3 & $D_2$\\
		\labelText{24C}{cl:4.0.0.6} &24 &  $[[4,0],[0,6] ]$ & 1.50 &  0 &  3 & $C_2$\\
		\labelText{25}{cl:5.0.0.5} &25 &  $[[5,0],[0,5] ]$ & 1.00 &  0 &  0 & $D_6$\\
		\labelText{27A}{cl:3.3.3.-6} &27 &  $[[3,3],[3,-6] ]$ & 1.00 &  1 &  0 & $D_6$\\
		\labelText{27B}{cl:3.0.0.9} &27 &  $[[3,0],[0,9] ]$ & 3.00 &  1 &  0 & $D_2$\\
		\labelText{28A}{cl:2.4.6.-2} &28 &  $[[2,4],[6,-2] ]$ & 1.00 &  0 &  3 & $C_6$\\
		\labelText{28B}{cl:4.0.0.7} &28 &  $[[4,0],[0,7] ]$ & 1.75 &  0 &  0 & $C_2$\\
		\labelText{30A}{cl:3.3.5.-5} &30 &  $[[3,3],[5,-5] ]$ & 1.04 &  0 &  1 & $D_2$\\
		\labelText{30B}{cl:5.0.0.6} &30 &  $[[5,0],[0,6] ]$ & 1.20 &  0 &  1 & $C_2$\\
		\labelText{32}{cl:2.4.6.-4} & 32 &  $[[2,4],[6,-4] ]$ & 1.00 &  0 &  3 & $D_2$\\
		\labelText{35A}{cl:1.5.5.-10} &35 &  $[[1,5],[5,-10] ]$ & 1.56 &  0 &  0 & $D_2$\\
		\labelText{35B}{cl:5.0.0.7} &35 &  $[[5,0],[0,7] ]$ & 1.40 &  0 &  0 & $C_2$\\
		\labelText{36}{cl:6.0.0.6} &36 &  $[[6,0],[0,6] ]$ & 1.00 &  1 &  3 & $D_6$\\
		\labelText{39}{cl:2.-7.5.2} &39 &  $[[2,-7],[5,2] ]$ & 1.00 &  1 &  0 & $C_6$\\
		\labelText{40A}{cl:3.-7.4.4} &40 &  $[[3,-7],[4,4] ]$ & 1.14 &  0 &  1 & $D_2$\\
		\labelText{40B}{cl:2.4.8.-4} &40 &  $[[2,4],[8,-4] ]$ & 1.31 &  0 &  3 & $C_2$\\
		\labelText{42A}{cl:3.-6.4.6} &42 &  $[[3,-6],[4,6] ]$ & 1.68 &  0 &  1 & $D_2$\\
		\labelText{42B}{cl:6.0.0.7} &42 &  $[[6,0],[0,7] ]$ & 1.17 &  0 &  0 & $C_2$\\
		\labelText{49}{cl:7.0.0.7} &49 &  $[[7,0],[0,7] ]$ & 1.00 &  0 &  0 & $D_6$\\
		\labelText{56}{cl:7.0.0.8} &56 &  $[[7,0],[0,8] ]$ & 1.14 &  0 &  1 & $C_2$\\
		\hline
		\hline
	\end{tabular}
	}
\end{table}
\newpage
\end{appendix}

\bibliographystyle{apsrev4-2.bst}

\end{document}